\documentclass[preprint,3p,times,authoryear]{elsarticle}

\usepackage{graphicx}
\usepackage{amsmath}
\usepackage{amssymb}
\usepackage{amsfonts}
\usepackage{mathrsfs}
\usepackage{adjustbox}
\usepackage{mdframed}
\usepackage[figuresright]{rotating}

\usepackage[final]{changes}

\usepackage{lipsum}

\usepackage{dutchcal}

\graphicspath{ {figures/} }

\usepackage{nomencl}
\usepackage{etoolbox}
\usepackage{wrapfig}
\usepackage{framed} 
\usepackage[skins,breakable, most]{tcolorbox}

\renewcommand*\nompreamble{\begin{multicols}{2}}
\usepackage{epsfig}
\usepackage{epstopdf} 
\usepackage{todonotes}
\usepackage{setspace}
\usepackage{siunitx}
\usepackage{tabularx}
\usepackage{longtable} 

\usepackage[utf8x]{inputenc} 
\usepackage{stfloats}

\usepackage{verbatim}

\usepackage[caption = false]{subfig}
\usepackage[hyphens]{url}
\usepackage{graphicx}
\usepackage{booktabs}

\usepackage{lmodern,textcomp}
\usepackage{rotating} 
\usepackage{multicol} 
\usepackage{multirow} 
\usepackage{mathtools, cuted}
\usepackage{lipsum}
\usepackage{tabularx} 
\newcolumntype{L}[1]{>{\raggedright\arraybackslash}p{#1}} 
\newcolumntype{C}[1]{>{\centering\arraybackslash}p{#1}} 
\newcolumntype{R}[1]{>{\raggedleft\arraybackslash}p{#1}} 
\usepackage{amssymb}
\usepackage{pifont}
\RequirePackage[ngerman=ngerman-x-latest]{hyphsubst}
\usepackage{microtype}
\usepackage[american]{babel} 
\usepackage{hyperref}
\usepackage{float}
\usepackage[normalem]{ulem}
\usepackage{colortbl}
\useunder{\uline}{\ul}{}
\usepackage{lscape}

\usepackage{amsmath, amssymb}

\usepackage{lineno}

\biboptions{square,comma,sort&compress}

\setcitestyle{numbers}
\journal{}
\makeatletter
\def\ps@pprintTitle{%
 \let\@oddhead\@empty
 \let\@evenhead\@empty
 \def\@oddfoot{}%
 \let\@evenfoot\@oddfoot}
\makeatother
\usepackage{longtable}  
\usepackage{tabulary}

\renewcommand{\nompreamble}{This list presents relevant symbols that are used within the body of this work. Most scalars will be denoted with cursive letters (mathcal style), function with gothic letters (mathfrak style), indices, graphs and most sets with roman letters and systems with cursive letters (mathscr style)}

\renewcommand\nomgroup[1]{%
  \item[\bfseries
  \ifstrequal{#1}{S}{Sets}{%
  \ifstrequal{#1}{F}{Functions}{%
  \ifstrequal{#1}{I}{Indices}{%
  \ifstrequal{#1}{M}{Vectors \& Matrices}{%
  \ifstrequal{#1}{G}{Graphs}{%
  \ifstrequal{#1}{Y}{Systems}{%
  \ifstrequal{#1}{C}{Scalars}{}}}}}}}%
]}
\makeglossary
\makenomenclature

\begin{document}

\begin{frontmatter}


\title{Stakeholder dynamics in residential solar energy adoption: findings from focus group discussions in Germany}



\author[DTU,IIRM]{Fabian Scheller}
\ead{fjosc@dtu.dk}
\author[IIRM]{Isabel Doser}
\author[IIRM]{Emily Schulte}
\author[IIRM]{Simon Johanning}
\author[ABE]{Russell McKenna}
\author[IIRM]{Thomas Bruckner}
\cortext[cor1]{Corresponding author}

\address[DTU]{Energy Economics and System Analysis, Division of Sustainability, Technical University of Denmark (DTU)}
\address[IIRM]{Institute for Infrastructure and Resources Management (IIRM), Leipzig University}
\address[ABE]{School of Engineering, University of Aberdeen}

\begin{abstract}
Although there is a clear indication that stages of residential decision making are characterized by their own stakeholders, activities, and outcomes, many studies on residential low-carbon technology adoption only implicitly address stage-specific dynamics. This paper explores stakeholder influences on residential photovoltaic adoption from a procedural perspective, so-called stakeholder dynamics. The major objective is the understanding of underlying mechanisms to better exploit the potential for residential photovoltaic uptake. Four focus groups have been conducted in close collaboration with the independent institute for social science research SINUS Markt- und Sozialforschung in East Germany. By applying a qualitative content analysis, major influence dynamics within three decision stages are synthesized with the help of egocentric network maps from the perspective of residential decision-makers. Results indicate that actors closest in terms of emotional and spatial proximity such as members of the social network represent the major influence on residential PV decision-making throughout the stages. Furthermore, decision-makers with a higher level of knowledge are more likely to move on to the subsequent stage. A shift from passive exposure to proactive search takes place through the process, but this shift is less pronounced among risk-averse decision-makers who continuously request proactive influences. The discussions revealed largely unexploited potential regarding the stakeholders local utilities and local governments who are perceived as independent, trustworthy and credible stakeholders. Public stakeholders must fulfill their responsibility in achieving climate goals by advising, assisting, and financing services for low-carbon technology adoption at the local level. Supporting community initiatives through political frameworks appears to be another promising step.

\end{abstract}


\end{frontmatter}
\section*{Highlights}
\begin{itemize}
\item Focus group discussions about stage-specific stakeholder influences on PV adoption.
\item Shift from passive exposure to proactive search for information during decision-making.
\item Influence of social network vanishes for less intimate stakeholders.
\item Desire for information from trustworthy and expert stakeholders increases.
\item Local utilities and authorities are trustworthy experts not exploiting their potential.
\end{itemize}



\section{Introduction}
\label{S:1}

\subsection{Social influences throughout the adoption decision-making process}
\label{S:1.1}
Despite extensive research on low-carbon technology (LCT) adoption in general (e.g., \cite{Curtin.2019, Geels.2018, Heiskanen.2017}) and residential photovoltaic adoption (PV) in particular (e.g., \cite{Alipour.2020, Galvin.2020, Mundaca.2020}), research explaining residential decision-making behaviour from a social-psychological perspective “focuses overwhelmingly upon individual consumers and under-appreciates the importance of interactions with other actors” \cite{Geels.2018}, resulting in a substantial gap regarding social dynamics driving adoption decisions \cite{Palm.2017}. It was no later than the 1920s that researchers found that innovation diffusion is a social process and communication structures, communication channels and the relations between sender and receiver influence what information is perceived and how it is interpreted \cite{Rogers.2003}. Yet, compared to the typically assessed individual level variables such as socio-demographic (e.g. \cite{Reames.2020}) and psychological factors (e.g. \cite{Sun.2020}), the intrapersonal nature of decision-making has received less attention, although without doubt, “the route that leads to one’s decision does not take place in a vacuum” \cite{Straub.2009}. Indeed, LCT adoption literature clearly indicates the dynamic nature of stakeholder influences throughout adoption processes (e.g., \cite{Rai.2016b,Palm.2017,Berardi.2013,Owen.2014}). Moreover, “social influence is often poorly theorized or simply absent from behaviour models” \cite{Axsen.2012}, reducing their usefulness to explain and predict adoption behaviour.

Drawing on social psychology literature, social influence is defined by French and Raven as “a change in the belief, attitude, or behaviour of a person […], which results from the action of another person” (\cite{french1959bases}, as cited in \cite{raven2008bases}). Since each stage of decision-making “is characterized by its own stakeholders, activities, and outcomes” \cite{Kamal.2011} and “the intensity of different stakeholders’ involvement in a particular phase may vary” \cite{Kamal.2011}, the term stakeholder dynamics describes the context- and time-dependent changing viewpoints, role involvements as well as degrees of social influence of stakeholders along the different decision-making stages of the low-carbon technology adoption process \cite{Postema.2010,Scheller.2020}. In fact, there are two levels of time-related dynamic origins: The first level is constituted by the procedural characteristic of innovation adoption and thus corresponds to the stages within the decision-making process \cite{Wilson.2018,Arts.2011}. The second origin is found on a higher level outside the adoption process, related to the rate of innovation diffusion \cite{Graziano.2015,Rode.2020}. The first dynamic can be captured by executing distinct stakeholder analyses for every process stage. Dynamics stemming from a higher level, however, can only be considered by conducting stakeholder analyses at different points in time. Considering the methodological possibilities and limitations of this paper, the focus lies on the former option. The investigation of stage-specific salience dynamics during the adoption process seems valuable to predict adoption outcomes more reliably \cite{Postema.2010}.

The lack of knowledge about stakeholder dynamics can potentially be attributed to several causes. First, influencing stakeholders are not always identically referred to in previous research (e.g. \cite{Rai.2016b} compared to \cite{Alipour.2020}). Second, only single stakeholders and their influence on decision-makers are investigated  (e.g. \cite{Owen.2014} assesses the influence of low carbon retrofit advisors and installers). Third, effects have mostly been investigated quantitatively \cite{Palm.2017}. Lastly, literature has shown that stakeholder circumstances, but also the credibility or rather the perceived stakeholder attributes, play an important role (e.g. likeable \cite{Palm.2017} or close \cite{Bale.2013} stakeholders are more influential), yet this research avenue has remained largely unexplored. 

\subsection{Research objective and methodology}
\label{S:1.2}
With regard to the outlined shortcomings of the existing literature, this paper qualitatively explores the influences of relevant stakeholders on residential PV adoption decisions throughout the decision-making process. PV was chosen due to the expected strategic importance for the future energy system as well as the increasing awareness of the technology in society (cf., \cite{IEA.2019}). For a systematic approach, we aim to answer the following research question: Which stakeholders are most influential in household decision-makers' choice to adopt residential PV? On the basis of the elaborated stakeholder dynamics from a residential perspective, policy recommendations to accelerate the uptake of residential PV shall be derived. 


To answer the research question raised and to overcome the gap of qualitative empirical research, focus groups have been identified as a suitable method due to their ability “to gain an in-depth understanding of social issues” \cite{Nyumba.2018} and their appropriateness to investigate group reactions to particular problems and processes in new fields of analysis \citep{GAILING2018355}. 
Therefore, they are helpful to investigate complex topics and to capture perceptions, opinions and feelings of people underlying their behaviour \cite{krueger2014focus}. To analyze the resulting data, the widespread approach of a structured content analysis is deemed suitable in this specific context \cite{mayring2014qualitative,SOVACOOL201812}. Elements of social network theory \cite{Wassermann.1994,Perry.2018,marin2011social} help to visualize both the stakeholder network surrounding a residential decision-maker and its dynamics. Instead of taking “a bird’s-eye view of social structure” to visualize a whole social network, egocentric networks capture only direct relations (ties) between alters surrounding a central actor termed ego \cite{marin2011social,Perry.2018}. In line with the research objective, the ego network visualization in this paper depicts relevant dyadic ties (representing influences) between a household decision-maker and alters (representing relevant stakeholders) during the PV decision-making process.


\section{Related work}
\label{S:2}

\subsection{Procedural decision-making}
\label{S:2.1}
The decision process preceding consumption behaviour is acknowledged to be of complex, intrapersonal nature \cite{Niamir.2018,Rogers.2003,Peattie.2010,Wilson.2007}, which is vividly shown by the richness of theorized explanatory variables, and the lack of a commonly agreed process \cite{Rogers.2003,Straub.2009,Wilson.2007,Wolske.2017}. Although the procedural nature is oftentimes only accounted for implicitly, several indications for different stages arise from literature. To determine how far advanced a decision unit is regarding the adoption process, knowledge about the technology \cite{Labay.1980}, engagement in activities to prepare behaviour \cite{Wolske.2017,Rai.2015b} or simply the decision units intents to perform the behaviour \cite{Sun.2020,Parkins.2018,Claudy.2013, MacPherson.2013, Ozaki.2011} are assessed. Only a few studies investigate the beginning of the decision process: In a study on PV adoption, Rai et al. \cite{Rai.2016b} ask for spark events, and Wilson et al. \cite{Wilson.2018} propose that property and household characteristics and conditions of domestic life primarily determine whether a decision-making unit begins to think about home refurbishments in the first place. Palm \cite{Palm.2017} suggests that in case of lacking awareness, visual impressions of local PV could play an important role in raising this awareness. Taken together, despite there being no consensus on how to clearly distinguish different stages, scholars seem to agree that with increasing closeness to adoption, considerations become more specific, concrete and context-dependent \cite{Wilson.2018,Arts.2011}.

\subsection{Stakeholder dynamics}
\label{S:2.2}
Influential stakeholders can roughly be distinguished into five groups \cite{Scheller.2020}: (1) social network including private persons, (2) stakeholders related to the energy system and PV related services, (3) government, authorities and other institutions, (4) others, including potentially influencing stakeholders related to e.g. the building sector or science and (5) the media.

The influence of the social network is investigated frequently under the umbrella term “peer effects". In the energy context, the term peer effects typically describe how individual behaviour is influenced by the behaviour of peers. Oftentimes, peer effects are studied by demonstrating a spatial clustering of a specific behaviour (for a detailed examination of peer effects see \cite{Wolske.2020}). Bollinger and Gillingham \cite{Bollinger.2012} and Rode and M\"uller \cite{Rode.2020} provide strong evidence for a cause-and-effect relationship between proximate prior PV installations and a household’s decision to adopt in the U.S. and Germany. Whereas according to Graziano \cite{Graziano.2019} the spatiotemporal effect is smaller for new PV systems in inner cities, Dharshing \cite{Dharshing.2017} finds ambiguous effects of settlement structure on PV adoption. While the passive peer effect (visibility of the panels) represents a substantial part of the entire peer effect according to \cite{Bollinger.2012, Rai.2013, Rode.2020}, Palm \cite{Palm.2017} as well as Mundaca and Samahita \cite{Mundaca.2020} find that the active peer effect (word-of-mouth) is more crucial. Mundaca and Samahita \cite{Mundaca.2020} additionally show that both seeing and hearing about PV systems from someone personally known to the respondent significantly increases the odds of having a higher likelihood of adoption, pointing towards stronger peer effects in case of combined physical closeness and likeability. Providing more evidence for the distinction of peers based on their relationship with the decision-maker, Palm \cite{Palm.2016} further finds that “neighbours were rarely perceived as influential in comparison to local persons to whom the respondents had a personal relationship". Both peer effects are found to decrease with distance \cite{Graziano.2015,Rode.2020} and to be stronger at early stages of PV diffusion due to higher uncertainty \cite{Rai.2013,Wolske.2017}. Whilst spatial proximity appears to be crucial for both, the effect of a mere visual impression, irrespective of who or what exactly is the owner of the visible PV system, and word-of-mouth, the effect of word-of-mouth appears to be governed by the closeness of the personal relationship between the previous adopter and the decision-maker. According to Rai et al. \cite{Rai.2016b}, a “conversation with friend/family/work” was crucial for only a minority of adopters for raising awareness in terms of a spark event. Yet for this particular adopter group, information received through their social network was more important in later stages of decision-making, than for groups who reported other spark events.
Overall, a considerable spatiotemporally and interpersonally dependent effect of previous adopters in the social network could be identified in the literature.

Concerning energy system stakeholders and PV related services, the current literature offers multiple indications: The PV manufacturer has only indirect influence through product-specific attributes like aesthetics and costs \cite{Curtius.2018}, and mostly functions as a wholesaler selling the product to PV providers \cite{Dewald.2012}. As the PV providers represent the interface between the manufacturer and the household decision-maker, they intuitively play a central role. This suppositionally important role is largely confirmed in literature \cite{Dewald.2012,Karakaya.2015b,Owen.2014,Rai.2016b}. Fabrizio and Hawn \cite{Fabrizio.2013} found that the number of PV installations increased more in cities where the presence of local installers was higher. While one study finds that such companies were perceived as a very competent information source later on in the decision-making \cite{Rai.2016b}, Palm and Eriksson \cite{Palm.2018b} report that households assessed information from PV providers as too technical. Berardi \cite{Berardi.2013} describes PV providers and energy advisors as influential. Despite their supposed and proven importance, both stakeholders are considered overlooked change agents \cite{Owen.2014}. Through strategic and proactive awareness campaigns, local utilities acted as an important stakeholder promoting local PV adoption in Sweden \cite{Palm.2016}. Similarly to the social network, literature stresses the reinforcing effect of spatial proximity on stakeholder influences.

Within the cluster government, authorities and other institutions, key influencing dynamics are attributed to local governments and authorities throughout the decision-making process by Bale et al. \cite{Bale.2013}. They are seen as a “trusted source of information” which “have local knowledge of the needs of their residents and communities” \cite{Bale.2013}. Consequently, poor local government engagement might even undermine political efforts at higher levels (state or federal) \cite{Berardi.2013}. Municipalities can be decisive role models at the awareness stage (e.g., installing LCTs such as PV on municipal buildings \cite{Kern.2005}) but also proactively supportive during PV planning (e.g., organizing central handling of grant applications \cite{Jager.2006}).

Also, broadly overlooked is the gatekeeping influence of building professionals and building contractors \cite{Zedan.2018,Curtius.2018}, whose strong influence potential is nowadays only used to a limited extent \cite{Davies.2011,Parag.2014}. Not surprisingly, De Groote et al. \cite{Groote.2016} found a statistically significant impact of house ownership status on residential PV adoption. Particularly in Germany, where national policies practically inhibited concepts of "Mieterstrom"\footnote{The landlord installs a PV system and provides electricity to the tenants}, the unsolved problem of split incentives \cite{Sutherland.1996} led to neglectable adoptions in the case of rented properties, pointing towards the powerful role of landlords \cite{Bale.2013,Richter.2013}.
Particularly in less mature markets, solar initiatives and civil society groups exert a significant influence on the decision \cite{Dewald.2012}.
Surprisingly, empirical social science research often finds direct effects of researchers on the studied samples \cite{Doci.2014,Schelly.2014,Sloot.2018}, suggesting that at least some household decision-makers turn to science to find reliable and up-to-date information.

Despite media being acknowledged to be of importance for disseminating mass advertisement and thus raising awareness for innovations in society in the Diffusion of Innovation theory \cite{Rogers.2003}, the overall importance of the media seems to be moderate in the case of residential PV systems \cite{Southwell.2014,Reeves.2017} with varying influence strengths in terms of media channels and process stages \cite{Rogers.2003,Rai.2016b}.

\section{Empirical research design}
\label{S:3}

\subsection{Focus group design}
\label{S:3.1}
\begin{figure}[ht]
 \centering
 \includegraphics[width=1\textwidth]{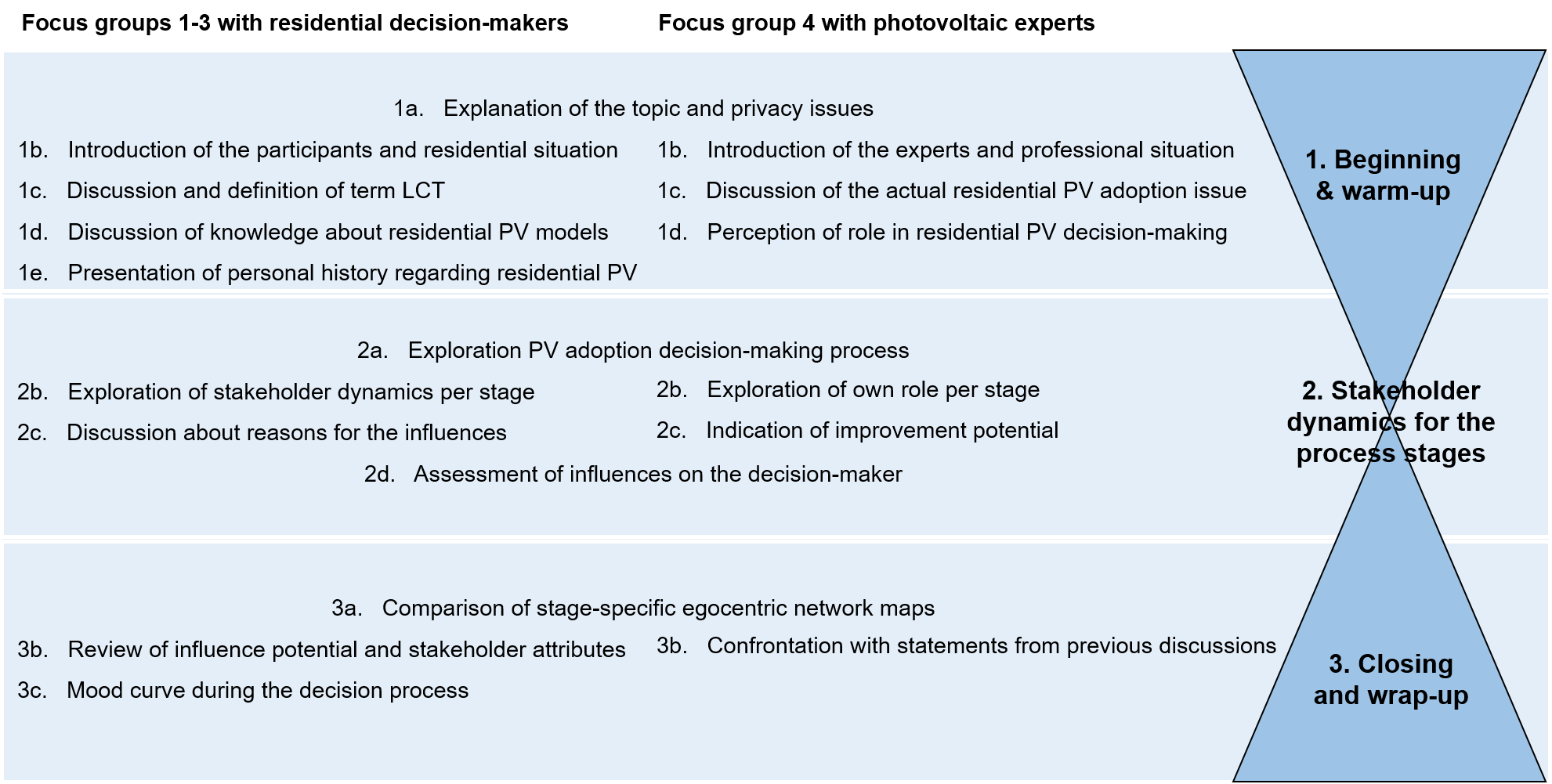}
 \caption{Brief description of the discussion guidelines for the focus group discussions with residential decision-makers (focus groups 1-3) and with photovoltaic experts (focus group 4) following the hourglass approach (own illustration).}
 \label{fig:houreglass}
\end{figure}

Following 
the “rule of thumb” of conducting three to five focus group sessions per research project \citep{Morgan.1997} to reach theoretical saturation (the point at which additional data do not provide new understanding and insights \citep{Glaser.1967,krueger2014focus}), four focus groups were conducted in close collaboration with SINUS Markt- und Sozialforschung GmbH (SINUS)\footnote{SINUS (https://www.sinus-institut.de/en/) is an independent institute operating in the field of social science research and consultancy. The institute offers a broad service portfolio for qualitative and quantitative research and is specialized in target group segmentation (Sinus-Milieus\textsuperscript{\textregistered}). The segmentation is based on a social milieu approach that groups milieus which could be described loosely as groups of like-minded people, using social status, lifestyles and basic values. This approach is a powerful means for moving beyond the traditional socio-demographic criteria.}. A strong focus on the evaluation of the conducted discussions as suggested by \cite{GAILING2018355} was taken into account with a full transcription of the focus groups for structured content analysis. Three focus groups were conducted with household decision-makers (focus groups 1, 2, 3), and one with a panel of PV experts who are influential stakeholders in the residential decision process (focus group 4) \citep{Scheller.2020}.
As recommended in the literature \cite{schulz2012quick}, structural guidelines were developed to support the moderator in keeping the discussion focused on the topics to be explored. The guidelines for focus groups 1-3 and focus group 4 are provided in Supplementary Material A and B. A visualization of the structure is provided in Figure \ref{fig:houreglass}. In line with Hennink \cite{hennink2013focus}, the guideline follows an hourglass design, starting with broad questions, followed by the key questions, and ending with again broader closing questions. 

To start with and to get a first impression of possible underlying causes, associations regarding LCT adoption in general and PV, in particular, were discussed. Subsequently, opinions about the decision-making process were obtained. To explore the decision process, a flip chart with a prepared graphical illustration of a three-staged adoption process (trigger, planning, realization) was shown to the participants, with only very rudimentary definitions provided if asked for. Thereafter, stakeholder influences were explored per stage and the moderator sketched egocentric networks on the flip chart concurrently. To guide the discussion, stakeholder categories from literature analysis by Scheller et al \cite{Scheller.2020} as presented in table \ref{tab:stakeholder} were used: social network, energy system and PV-related services, government/authorities and other institutions, media and others. Furthermore, additional questions concerning stakeholder salience, kind of influence, expectations and trust were posed. In contrast to focus groups 1-3, during which participants were asked to describe their circumstances and experiences as a (potential) PV adopter, the expert panel was asked to discuss their perceived roles within the stages, communication channels used and reflect upon optimization potentials.

\subsection{Sample selection and practical implementation}
\label{S:3.2}
For each focus group, in line with recommendations \cite{hennink2013focus,krueger2014focus}, a group size of six was aimed at. 
The participant recruitment for focus groups 1-3 was carried out by the local market research service provider IM Teststudio Leipzig. For the 18 household representatives, a screener was developed to identify suitable participants. Participants (1) must have experienced at least two out of the three stages of the adoption process, and (2) must be involved in energy-related decision-making processes in their household. Each of the focus groups 1-3 consisted of representatives of similar milieus. Since there seems to be a connection between lifestyles such as the Sinus-Milieus\textsuperscript{\textregistered} and energy-related adoption decision \citep{Groger.2011,glockner2010lohas}, the following milieus were chosen and clustered by similarity: (1) Modern Mainstreamers, Adaptive Navigators, Social Ecologicals; (2) Established, Liberal Intellectuals; (3) Postmodern Performers, Cosmopolitan Avantgardes. Brief profiles of the seven included milieus are given in the Supplementary Material D. 

Despite a potential bias issue, as residents of lower social status make up a huge portion of the population and potential PV adopters, the decision for homogeneous milieus in and across the focus groups was based on the following two reasons. First, due to the relatively high financial effort related to PV technology \cite{Bashiri.2018,Jager.2006}, milieus with a relatively low social status or with a small percentage of property owners (Traditionals, Precarious, Hedonists) were not considered. Relevant stakeholder dynamics across the different decision-making stages are only demonstrable if the participants can take an adoption decision and have already considered or even implemented it. Similar procedures are found in other empirical works related to PV \cite{Claudy.2013,Rai.2015,Parkins.2018,aziz2017factors}. The investigations on water use innovations by Schwarz and Ernst \cite{Schwarz.2007,Schwarz.2009}, on green electricity consumption by Ernst and Briegel \cite{ERNST2017183}, and on sustainable heating system adoption by Gr{\"o}ger et al. \cite{Groger.2011} also support the selection since most of the innovation characteristics of higher-income milieus are considered to be more positive than the others. Second, the homogeneity within focus groups settings might facilitate a more robust conversation which in turn might lead to a better understanding of particular themes. Participants should be able to construct responses based on what they hear from fellow participants in focus groups \cite{SMITH2018307}. In this regard, the participant selection of the focus groups goes in line with the experiences of Smith et al. \cite{SMITH2018307} and the suggestions of Hennink \cite{hennink2013focus} and Smithson \cite{Smithson.2000}. 

An overview of the sample is outlined in Table \ref{tab:sample_overview}. All focus groups had a well-balanced gender distribution. In total, twelve participants were house owners. Five participants had already successfully adopted rooftop PV systems, nine participants were still within the rooftop PV decision-making process, two participants with tenant status were only interested in the adoption of PV modules for balconies and one participant had to reject PV adoption due to tree shading. It came out that one participant had only invested money in a closed-end fund for PV technology as a monetary investment and therefore does not match the research focus; not all of his or her contribution during the focus groups are therefore fully usable. In terms of focus group 4, the participants were recruited by SINUS in cooperation with the research team. Six PV experts with different backgrounds from Leipzig and the Leipzig area were invited.

\begin{table}
\footnotesize
\renewcommand{\arraystretch}{1.2}
\setlength{\tabcolsep}{5pt}
    \caption{Overview of the day, time, and sample of each of the conducted focus groups 1-4 (own composition).}
    \label{tab:sample_overview}
\begin{tabularx}{\textwidth}{ p{0.03\textwidth} p{0.07\textwidth} p{0.08\textwidth} p{0.15\textwidth} p{0.15\textwidth} p{0.13\textwidth} p{0.2\textwidth}}
\toprule
Focus group & Date & Time & Participants & Housing situation & PV adoption progress & Lifestyle/ Profession \\ 
\toprule
1 & 9th Sep 2019 & 5-7 p.m. & Decision-makers (3 female, 3 male);  & 
3 single-family houses, 3 tenants in apartment buildings &
5 interest/early planning stage, 1 PV adoption
& Modern Mainstreamers, Adaptive Navigators, Social Ecologicals
  \\
\hline
2 & 9th Sep 2019 & 7.30-9.30 p.m.  & Decision-makers (3 female, 3 male);  &  4 single-family houses, 1 tenant in apartment building, 1 condominium in a multiple dwelling
& 4 interest/early planning stage, 1 PV adoption, 1 PV rejection (shading)
& Established, Liberal Intellectuals
 \\
\hline
3 & 10th Sep 2019 & 5-7 p.m. & Decision-makers (2 female, 4 male); & 3 single-family houses, 1 duplex house, 1 terraced house, 1 condominium in a multiple dwelling
 &
3 interest/early planning stage, 3 PV adoptions  & Performers, Cosmopolitan, Avant-gardes
 \\
 \hline
4 & 18th Sep 19 & 6-8 p.m.  & PV experts from Leipzig or Saxony (2 female, 4 male) &  -
 & -

 & Private energy consultancies, local energy provider, consumer advice centre, the municipal environmental authority
 \\
\bottomrule
\end{tabularx}

\end{table}
 
The sessions (approx. 120 minutes) were conducted in a professional environment in Leipzig, Germany (IM Teststudio) between the 9\textsuperscript{th} and 18\textsuperscript{th} of September 2019. Whilst the research team observed the discussion behind a mirrored wall, a senior consultant of the field partner conducted the focus groups. With the permission of participants, focus groups 1, 2 and 3 were video- and audiotaped. Focus group 4 was only audiotaped. All discussions have been fully transcribed in German and were used by the research team for analysis. 

\subsection{Analytical assessment of the focus group discussions}
\label{S:3.3}
The transcripts were analysed following a structured qualitative content analysis developed by Mayring \cite{mayring2014qualitative}, whose objective is “to filter out particular aspects of the material, to give a cross-section through the material according to pre-determined ordering criteria, or to assess the material according to certain criteria” \cite{mayring2014qualitative}. Since focus groups are explicitly mentioned by Mayring \cite{mayring2014qualitative} as a conceivable source of data and additionally, the data are available in the form of written transcripts, this method seems appropriate. The desired content is extracted by means of a category system which is not only the core instrument of analysis but at the same time operationalizes the analysis objectives \cite{mayring2014qualitative}. 
Despite the fact that the categories are literature-based, the qualitative content analysis follows an inductive and interpretive methodology as suggested by Sovacool et al. \cite{SOVACOOL201812}. In this regard, the structured analysis of focus groups is seen as a useful research method when there is a theoretical framework \cite{SOVACOOL201812}.

In the first step, the transcripts were screened regarding statements about the conceptual understanding of the underlying research framework in which the stakeholder dynamics are embedded. The understanding of the participants regarding the term LCT (Section \ref{S:4.1}), as well as associations with PV (Section \ref{S:4.2}), are investigated. The exploration of the procedural PV decision-making is presented in a discursive way in Section \ref{S:4.3}, and a decision process (Figure \ref{fig:decisionprocess}) for the further analysis regarding the stakeholder dynamics is derived. The obtained results are underpinned by a list of characteristic quotes extracted from the transcripts which are outlined in the original language German and the translated language English in Supplementary Material F-H.

In the second step, stakeholder dynamics were analysed using the category system shown in Table \ref{tab:category_system}. While questions 1 and 2 are built upon nominal categories derived from the literature-based stakeholder analysis, question 3 requires ordinal categories. Three assessments and three respective variables were used to collect the information needed to develop ego network maps including stakeholders, attributes, and strength of influence per stage. Initially, colour codes were assigned to all textual elements of the transcripts indicating any kind of stakeholder influence unless one participant brought up the very same content several times; in this case, the perceived influence was only coded once. When a stakeholder was not made explicit (e.g. utility company without specifying local or national scale), the text passage was assigned to all respective potential stakeholders. The analysis has been conducted by two independent researchers to enhance the reliability of the results as requested by \cite{mayring2014qualitative}. Thereafter, the coded parts were transferred to a result table in an external document, categorized by process stage and by stakeholder, and analysed regarding relevant content for assessment questions 2 and 3. To assess influence strength more reliably, qualitative results were supplemented with frequency data, as suggested by Onwuegbuzie et al. \cite{Onwuegbuzie.2009}. The total number of participants from focus groups 1-3 that stated to have been influenced by a particular stakeholder in a particular stage was counted, as well as the number of focus groups in which the stakeholder was recognized as influential. The frequency data is provided in Tables \ref{focus groupsDs_1}, \ref{focus groupsDs_2}, and \ref{focus groupsDs_3}. Influence strength (strong, medium, small) was then assessed by taking into account the textual content, the assignment of attributes, and the frequency data. While the results per stakeholder are outlined in Section \ref{S:4.4}, the results per decision-stage are outlined in Section \ref{S:4.5}. Mixing qualitative and quantitative methods allows the development of a more detailed view of the meaning of a phenomenon for individuals \cite{EDLING2018331}.

\begin{table}
\footnotesize
\renewcommand{\arraystretch}{1.2}
\setlength{\tabcolsep}{5pt}
    \caption{Category system for data analysis of focus groups. The assessment questions are formulated for each decision-making stage respectively (own composition).}
    \label{tab:category_system}
\begin{tabularx}{\textwidth}{ p{0.025\textwidth} p{0.45\textwidth} p{0.2\textwidth} p{0.2\textwidth}}
\toprule
& Assessment question & Variable & Value  \\ 
\toprule
1 & By which stakeholder(s) is the Ego (household decision-maker) influenced?  & 
Influential stakeholder
 & Nominal: \newline
List of stakeholders from literature analysis
  \\
\hline
2 & Which attributes are subject to the interaction with stakeholder X? & Attribute perception & Nominal: \newline
Expertness, Trustworthiness, Power, Likeability, Closeness
 \\
\hline
3 & How strong is the influence of stakeholder X in the different stages of the decision-making process? & 
Influence strength &
Ordinal: \newline
Strong, middle, small, none \newline
Frequency count
 \\
\bottomrule
\end{tabularx}

\end{table}

\section{Results of the focus group discussions}
\label{S:4}

\subsection{Understanding of residential low-carbon technologies}
\label{S:4.1}
The participants of the focus groups had a similar understanding of focal terms\footnote{A list of characteristic quotes of the focus groups that support the line of reasoning in the following is provided in Supplementary Material F-H. This list includes the most representative original statements of the transcribed focus groups in German and the corresponding translated statements in English of the participants classified by concept, stage, and stakeholder in order to illustrate and underpin the written results on conceptual perception and stakeholder dynamics.}. The open question regarding their understanding of LCTs confirmed the predetermined definition\footnote{We follow the definition elaborated in Scheller et al. \cite{Scheller.2020}: Low-carbon technologies represent innovative consumer-side technologies whose adoption requires a deliberate decision-process that have the potential to reduce residential GHG emissions by decreasing overall demand for carbon-intensive energy from fossil fuels through improving energy efficiency and/or generating no- or low-carbon energy.}. According to the participants, LCTs are complementary to standard technologies, A+++ technologies, energy-saving devices, or environmentally friendly technologies. They consume less energy, increase the degree of efficiency, and utilize renewable energy sources. According to this, LCTs support climate protection, help to reduce carbon emissions, increase energy autonomy, and reduce the consumption level. Nevertheless, there were individual critical voices regarding the valuable contribution of LCTs when considering the whole product life cycle. Some participants also related LCTs to better filtration techniques which reduce exhaust gas pollution\footnote{This might be related to the German translation used for LCT (“emissionsreduzierende Technologien” (emission-reducing technologies)) which is not as clear as the English term low-carbon technology in this context.}. Example technologies that were reported by the participants include heat pumps, geothermal heating systems, combined heat and power, photovoltaic panels, solar thermal technologies, fuel cells, wall and attic insulation, electric vehicles, hybrid vehicles, efficient fridges, smart washing machines, low-energy and LED bulbs). While the majority across all focus groups named cost reduction as major adoption reason, technology affinity, ecological claims, or social norms were also mentioned. 

\subsection{Associations regarding photovoltaic and storage options}
\label{S:4.2}
The associations of the participants regarding PV technology covered economic, environmental, societal, and technical aspects. Both large- and small-scale PV installations were named and critically reviewed; there was no generally accepted view among the participants regarding the best sustainable option. The associations ranged from fascination and positive feelings over a rather neutral stance to even negative and bad associations. While the attitude towards the technology as such was in general positive, negative associations were mainly related to large-scale PV installations. It was stated that “[i]t is a shame to cover the wastelands with photovoltaic panels” and “to cover whole nature with PV panels is still sometimes strange”. Whereas “[a] roof, after all, is an unused area [...]. And installing PV panels on these roofs seems actually reasonable”. In contrast, another participant stated that “[s]uch a solar field is actually something good and meaningful, and that everybody has small-scale systems on their roofs is from my point of view rather strange”. Moreover, various participants considered storage options in conjunction with PV systems as a bundle. Participants perceived that PV systems were heavily advertised a few years ago, whereas today the focus is on storage options. 


\subsection{Exploration of the decision-making process}
\label{S:4.3}
The three ambiguous decision-making stages raised (trigger; planning; realization) during the focus groups were discussed lively by the participants. This helped to derive a decision-making process characterized by stages whose beginning is marked by a “Moment of Truth" (MOT)\footnote{MOTs are a concept used in marketing, describing moments where the inner attitude of customers concerning a product changes. In the context of this research paper, they represent points in time during the decision process after which the decision-makers actions and information needs change, represented by the respective subsequent process stage.}. 

All participants of focus groups 1-3 agreed that the first stage of their decision-making process was a period during which they became aware of PV technology. They described it as the stage of first impressions and inspiration, typical events are related to passive exposure in their environment, random web browsing and unintended chats about PV technology. Younger participants indicated that they have grown up with PV, whereas older participants told memories about their first “contact” with PV. Participants who have already adopted described their mood at the (end of the) first stage as being excited and curious. The majority of the expert panel also related the first stage to “awareness-raising for the topic”, “[w]hen the consumer develops the idea for the first time” and the attitude “[t]his could be something for me”.

At the stage of initial awareness, “a smooth transition” to the next stage, which was characterized by an existing interest in PV, was described. A participant stated that he “suddenly realized the option” and that “[t]his was the moment when [he] thought about getting more involved with the topic and to dig a little bit deeper”. “You have to start at some point for real and not always somehow [...] otherwise you will never do it”. Whilst some described the stage as a period of targeted research to increase knowledge, driven by experiences of previous adopters, as well as data and facts, for many, the planning stage involved not only the mere interest in the adoption but already a concrete intention: “When you say it is not only interesting, but I want to do it”. Furthermore, participants described it as the stage when “it becomes more concrete" and “you start to ask for quotes”. The latter understanding also emerged in focus group 4. Experts related this stage to a done deal and “the beginning of the investment because the planning process already entails costs”. Another expert stated: “The planning stage begins when I am in touch with an energy consultant or engineering office, perhaps I have already spent some money or asked for an assessment which entails costs. I have already taken my decision that I would like to realize it. However, it can still happen that the project fails due to certain conditions”.

As the understanding of the planning stage strongly deviated among participants, it was not surprising that the consecutive realization stage was also interpreted differently. Some participants stated that the implementation stage begins with the adoption decision made (“when I have signed the contract”) and contains "only the technical realization”, others and most of the experts interpreted the implementation stage as less advanced and evaluated it as object-specific engagement and activities that lead to a final decision, including e.g. explicit proof of feasibility, comparison of offers and selection of the final suppliers. One participant suggested the selection of a concrete product as the beginning of the stage, and pointed out that “it is still possible that I reject the adoption”. In line with this understanding, another participant located the event of contract signing rather to a point in time “[…] when [he or she is] already for a long time at the implementation stage”. The implementation stage was described positively throughout - somewhat less exciting than the trigger stage, but more relaxed, as adopters felt happy that most of the work related to the adoption of PV would be over soon.

\begin{figure}[ht]
 \centering
 \includegraphics[width=1\textwidth]{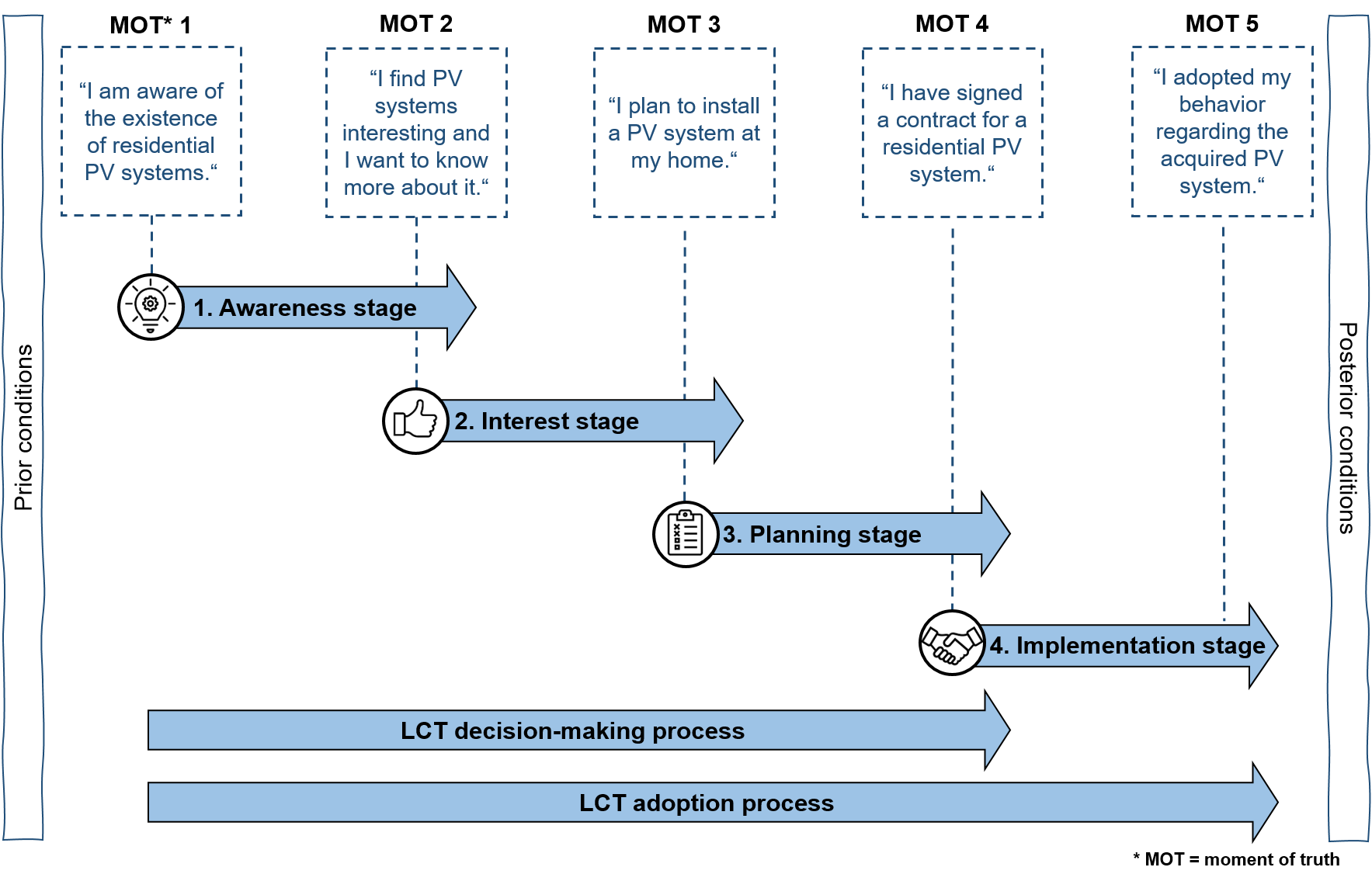}
 \caption{Process stages of residential low-carbon technology decision-making with particular attention to the behavioural transitions (MOT = moment of truth). According to the objective of this paper, the three stages of the decision-making process (awareness, interest, planning) are in focus (based on \citep{Scheller.2020}).}
 \label{fig:decisionprocess}
\end{figure}

Although the participants did not always agree on the beginning and content of the proposed ambiguous stages, a relatively clear picture of a decision-process with three subsequent stages emerged. Since nonetheless, a strict separation of the process stages is not realistic, overlapping stages are assumed. The process is shown in Figure \ref{fig:decisionprocess}. The first stage begins with initial awareness of the product and is characterized by passive exposure and a lack of intrinsic motivation to engage with the topic. This phase will be referred to as the awareness stage. At a certain point in time, individuals eventually perceive the product as interesting and begin to actively seek and evaluate information from various sources. They enter the interest stage. If the information search led to a sufficiently positive attitude, interest becomes intention, and the individual starts to collect object-specific information, including e.g. contacting a PV provider or energy advisor, or using an online feasibility calculation tool. Both interest and planning stages can cause an individual to discontinue the process, for example, negative news (interest stage) negatively affect intention, or lack of financial funds (planning stage) inhibit adoption. The decision-making process ends when an irreversible decision has been made - which is the case when a contract got signed. This doesn't conclude the LCT adoption process, however, as the posterior conditions are only reached after the technology has been installed, and the individuals' behaviour has been adjusted to the new situation. Therefore, the implementation stage is not part of the LCT decision-making process but of the adoption process.

\subsection{Perception of influences in terms of different stakeholders}
\label{S:4.4}
The results of the focus groups are subsequently presented within the defined stakeholder categories: Social network, energy system and PV-related services, government, authorities and other institutions, others, and media.

\subsubsection{Social network}
\label{S:4.4.1}

First of all, neighbours were perceived as an important stakeholder. They are seen as a trustworthy source of advice and existing PV systems as demonstration objects to reduce uncertainties and reinforce the adoption idea, especially at the beginning of the decision-making process. It was mentioned several times that participants asked them about common pitfalls to ensure trouble-free adoption. Several participants indicated that active peer effects in the form of informal chats among neighbours emerged from passive peer effects. It is assumed that they took place as a result of likeability since one participant indicated that “it is always good to discuss over a glass of wine or beer.” This assumption is further strengthened by a more suspicious voice as follows: “[…] when you move to a new neighbourhood like we did, [the neighbour] won’t tell us that he regrets his decision.” 
Another participant stated that one neighbour adopted PV a few years ago and today, half of the neighbourhood has installed PV systems. Furthermore, one participant (who didn’t have any previous adopters in his or her neighbourhood) even admitted that “if everybody had it, for sure, I would have dived more intensively into the topic.” Another participant supported this feeling as he or she reported that “I saw it in the neighbourhood […] and then I felt the urge to have it myself”.

Similar patterns were seen for further stakeholders within the social network. The partner is highly influential due to mutual decision-making, as “you have to convince your partner or vice versa. The others rather perform an advisory function.” Family and friends or acquaintances were mainly influential through active peer effects and also emotional support. The influence of children was for instance brought up by an expert who perceived that school projects about the topic lead to discussions and therefore the presence of the topic in families. Friends are seen as more trustworthy than neighbours; they are expected to be honest about negative aspects with one participant stating that “they would tell me more likely about drawbacks.” Friends with technical expertise were particularly influential at later stages (“our electrician, he’s my best friend”). Co-workers were mentioned slightly less often. One participant, nevertheless, evaluated the situation in a similar way to neighbours although “[…] you don’t see that there is a PV system on the roof, instead there are chats.” In contrast, participants who stated that any of these stakeholders was relevant also stated that their social network was not interested in such topics or didn’t have access to any previous adopters. The influence of the social network is further strengthened by passive peer effects from other people, especially previous adopters. In addition to the general environment, especially farmers seemed to have a huge impact on awareness-raising. One participant also evaluated members of online forums as important regarding past PV adoption experiences which was seen critically by another participant in terms of source credibility since “every amateur could write something which is not necessarily correct”.

\subsubsection{Energy system and PV-related services}
\label{S:4.4.2}
The function ascribed to local utilities was mainly seen limited to technical concerns. The interaction was mentioned to clarify if a PV installation can technically be realized, as well as later in the process shortly before adoption, where participants brought up the future contractual relationship regarding feeding-in. One participant explicitly pointed out that the local utility plays a minor role at early stages because at this state the decision-maker is not yet having concrete questions. “Anyway, the utility company is obliged to feed in the electricity and buy it from you." Compared to national utilities, participants stated that local ones are more influential and slightly indicated higher trustworthiness towards them. Interestingly, an expert doubted that decision-makers would attribute a higher degree of trust(worthiness) to local utilities in comparison to national ones. This suggests that local utilities are not aware of their potentially more favourable position, at least compared to national utilities. Moreover, another expert explicitly supports the latter assumption: “I would categorize the public local utility always just as an option or provider among many others. And not in a particular position…”. 
The non-existent influence apart from the mentioned technical concerns is especially interesting because the local utility in Leipzig offers turnkey PV systems for households. An expert stated in this context that the awareness about the product offered by the utility has not yet been raised. At the same time, participants indicated that the local utility could be more present and proactively inform citizens. Apparently, the local utility is seen as a legitimate and trustworthy stakeholder who does not yet touch upon its potential.

According to the focus groups, PV manufacturers are of minor importance. A minority of the participants indicated manufacturers as a source of information. At the awareness stage, there was a consent that they are not influential at all but some participants stated that it could be valuable for them to advertise more. One participant saw manufacturers influential at the interest and early planning stage as he or she investigated the current economic situation of the companies: “if I get ten years manufacturer warranty and after two years, […] the company is bankrupt, I cannot benefit from [it]”. 

In contrast, PV providers are recognized as one of the most important stakeholders. The contact was proactively established by the decision-makers (e.g. online and trade fairs). Participants see providers as the main source of information. Furthermore, they are seen also as competent and knowledgeable regarding funding opportunities and bureaucratic requirements. On the one hand, they were seen as experts who “are fully trust[ed] at the [planning and] implementation stage” as a result of lacking professional expertise of decision-makers. On the other hand, it is perceived that the provider “will not necessarily always act in the customer’s best interest” due to commercial intentions. Additionally, participants indicated difficulties in finding available providers (most importantly regarding craftsmen for the installation). This issue is especially true for apartment buildings: “We are doing our best to find someone, but the companies don’t answer anymore. […] Then you contacted four or five companies but it’s the same situation.” The influence of PV providers in terms of apartment buildings is therefore seen as highly hindering whereas their influence regarding single-family houses is positive and strong due to the variety of functions carried out, as well as the partly attributed expertness and trustworthiness.

None of the participants had involved energy advisors from the business sector in the decision-making process whereas advisors from NPOs\footnote{Verbraucherschutzzentrale and Industrie- und Handelskammer (Chamber of Industry and Commerce)} were contacted by two participants primarily at the interest stage and secondary at the planning stage. The minor role of advisors is supported by an expert who believes that enough advice is provided by local PV providers. Furthermore, distrust against commercial energy advisors was revealed: “This must be expensive. Who knows what they offer you.” The distrust is primarily rooted in the assumption that energy advisors are not independent actors. An expert partly confirmed this negative perception of decision-makers: “I had experiences with energy advisors who were responsible for electricity and gas […] they just name themselves advisors. They are salesmen.” In contrast, participants agreed that independence is given regarding consultancy from NPOs. Interestingly, the majority of decision-makers were not aware of energy advisory services offered by NPOs which is strongly confirmed by the panel of experts. Yet, higher involvement of such organizations was claimed, and one participant even suggested the German Consumer Organization as a desirable advisor, obviously lacking knowledge of the existing advisory service. The awareness was even lower regarding energy agencies, with participants not even knowing the term at all and asked what energy agency means. They suggested that energy agencies should engage more in external communication and highlight their competencies as independent experts. 

\subsubsection{Government, authorities and other institutions}
\label{S:4.4.3}
The influence of funding bodies and financial institutions was emphasized regarding subsidies as well as financing of the investment. Funding bodies were seen as enablers. While some participants criticized the “passive” position of funding bodies, others regarded proactive information procurement of household decision-makers as self-evident. Active information dissemination regarding subsidies is still seen as an appropriate means to trigger PV adoption. Several participants felt a lack of transparency due to the complex and rapidly changing funding landscape. Furthermore, the online provision of information was known by the participants, but personal advice was nevertheless seen as necessary. Two experts stated that the complexity can hardly be overcome through online-only information. A potentially inhibiting influence was attributed to them, as banks “can also complicate the procedure.” Furthermore, one participant wished “that I can approach my local bank regarding funding opportunities – […] That you do not have to contact 1,000 institutions.” All in all, both stakeholders prepare a ground for decision-making and further trigger decision-makers’ motivation in terms of profitability of the adoption, however with a discouraging effect due to uncertain and complex circumstances and intensified through lacking knowledge regarding points of contact and offers.

The recognized influence of local governments was mainly the power over granting building permits. In this context, the bureaucratic burden was criticized. Only one participant mentioned that he or she proactively involved the local government: “I even showed them my concrete quotes at the Technisches Rathaus [(Technical City Hall)] [...]. Including a map, the house, construction plan and they told me which quote makes sense.” This implies two assumptions. First, the participant must have had a lack of trust in PV providers and second, this trust gap could be overcome with the help of the local administration. In contrast, the awareness of this option must have been low because several participants stated that they would appreciate an independent consultant, optimally employed by the municipality and thus without commercial interest. In general, many participants indicated that they perceive a lack of information and support of local governments while clearly attributing responsibility regarding more proactive action in LCT adoption. One participant brought up that the municipality would be perceived as a role model if PV systems were for instance installed at municipal buildings and at the same time increase the awareness. 

The role of federal or non-local governments is acknowledged to set the framework for PV adoption and therefore to be highly influential although no direct interaction takes place. Regarding the awareness stage, it was desired to have an “impetus from politics and public institutions, because they decided to abandon nuclear energy. Therefore, they have to offer a suitable solution”. Moreover, the influence was assessed to be mostly of a negative nature. Criticism that the political apparatus lacks reliability emerged since the framework was described as rapidly changing. Experts criticized lacking monetary incentives concerning the current profitability of PV systems and too complicated framing conditions. Feed-in tariffs were generally perceived to positively affect adoption while their ongoing decrease was rated negatively. However, the decrease caused pressure to speed up adoption.

\subsubsection{Others}
\label{S:4.4.4}
Several participants who were engaged in housing construction indicated that building professionals were reliable and trustworthy influencing stakeholders at all stages. “The building contractor gave us advice regarding the construction of our house and also the neighbours. Almost all houses in our neighbourhood have PV systems on their roofs.” Another participant was strongly influenced by the architect which was, however, simultaneously a friend of the participants’ family. Thereby, the trust-based relationship was emphasized. In general, building professionals can objectively inform decision-makers during housing construction about LCT adoptions. 

Landlords were perceived as powerful regarding decision-making but in an inhibiting way. Most participants who were tenants did not even consider approaching their landlord in terms of PV installation as they assumed lacking willingness to engage in this topic: “Their priorities are clearly new constructions. And there, eventually PV systems are installed if it is seen […] profitable [...]”. Instead, more active engagement of both private owners and housing associations is claimed and responsibility regarding the commitment to sustainable energy is attributed.

Several participants pointed to the high relevance of knowledge from the scientific community, especially prevalent regarding the generation of knowledge. Both expertness and trustworthiness were attributed to this stakeholder while constituting a strong but indirect influence transferred via different media channels such as online sources or scientific journals. Some participants stated to trust most in scientists: “I don’t want to be naive but when scientists tell me the best model and how to do something, I do it this way […]. I am not a PV scientist.” Since scientific contributions are seen as a means to mitigate ambiguous information from different sources, the desire for more communication from scientist to the public in an understandable way was reported.

\subsubsection{Media}
\label{S:4.4.5}
Different media channels are perceived to exert influence at different stages. Participants indicated that a scientific character of all kinds of media content is important for credibility and thus influence strength. Distrust was, for instance, an issue regarding advertisement. The influence is further narrowed to being “at most a trigger to deal in greater detail with this topic.” Additionally, the likelihood of exposure to advertisement has decreased over time (“At the peak of PV, there were commercials omnipresent on [different television channels]…they have all disappeared”.) Nevertheless, one participant reported that his interest was sparked recently by a random exposure to an online advertisement. While mass media such as magazines, TV and films were recognized primarily as influential in terms of awareness, scientific journals served as a source for technical aspects and economic profitability assessment of the investment. 

The perceptions regarding social media were divided: older participants stated that they were not influential at all and that “I am not sure if there is something on these platforms worth the time spent”, whereas younger participants rated social media as very important. This was supported by an expert who assumed a low influence of social media on people in their mid-fifties and older. Also, unspecified online sources constitute a major influence, especially regarding proactive information gathering about how PV adoption works in general, technological aspects, availability of local PV providers and funding opportunities. However, online sources were also regarded critically: “The internet is boon and bane at the same time. I’ve learnt to never only trust information from sources making profit of it, like manufacturers”, and further, “[y]ou really have to diligently search for [objective information].” In this context, science also seems to drive source credibility since the level of trust in terms of information was linked to content being backed up by scientific studies. Furthermore, media in general are seen as a vehicle to reach information from the research community for instance regarding technical developments and improvements. To sum up, there is a shift from random influences at the awareness stage to targeted, proactive research at subsequent stages, accompanied by changing salient channels which facilitate both spark events and information procurement. 

\subsection{Perception of stakeholder dynamics with respect to the decision-making stages}
\label{S:4.5}
\begin{figure}
  \centering
  \includegraphics[width=1\linewidth]{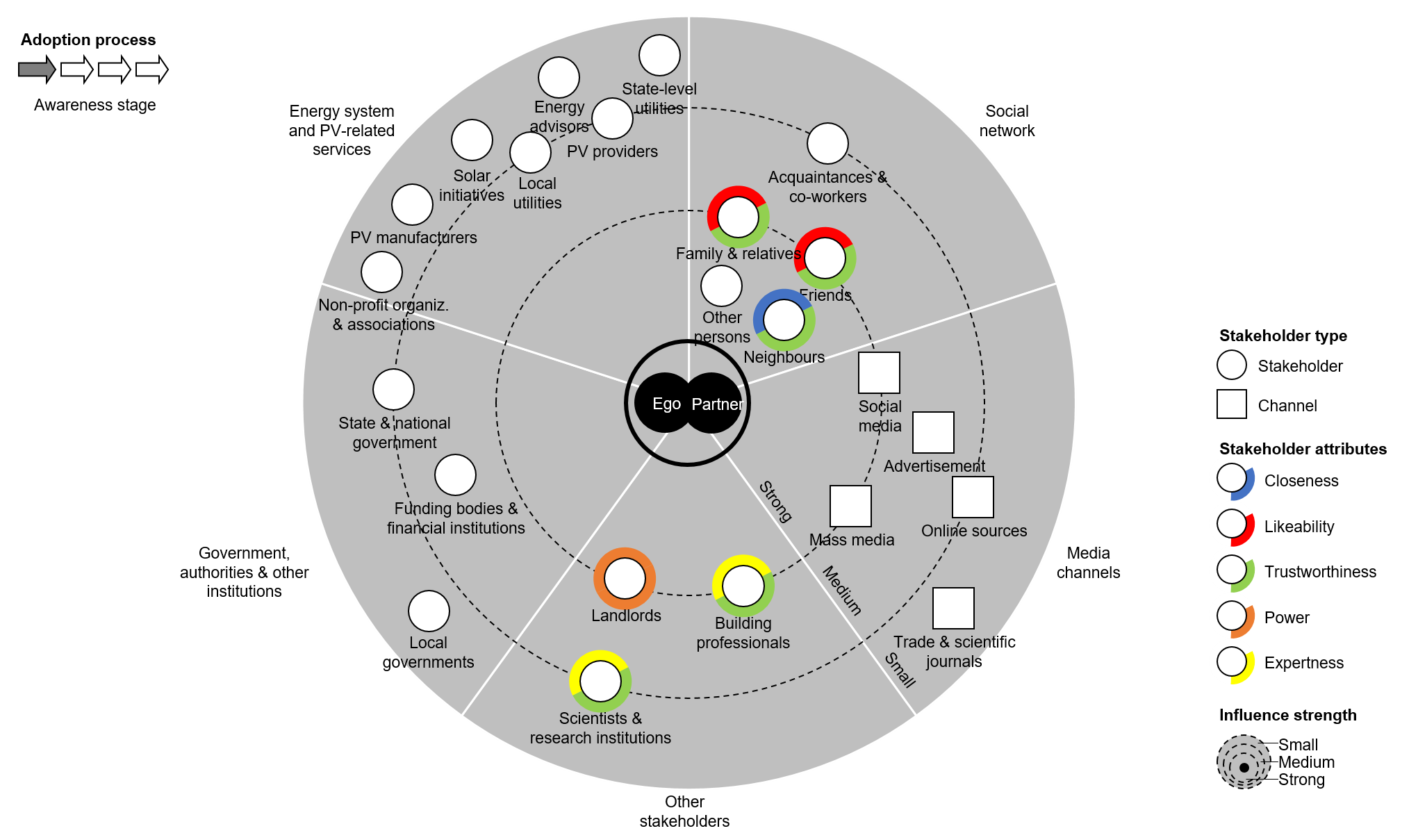}
  \caption{Illustrative overview of egocentric networks of the stakeholder dynamics at the awareness stage. The results show the stakeholder dynamics of the focus group discussions in terms of residential PV adoption from the perspective of the household decision-maker (own illustration).}
  \label{fig:EgocentricNetworksAwareness}
 \end{figure}

\begin{figure}
  \centering
  \includegraphics[width=1\linewidth]{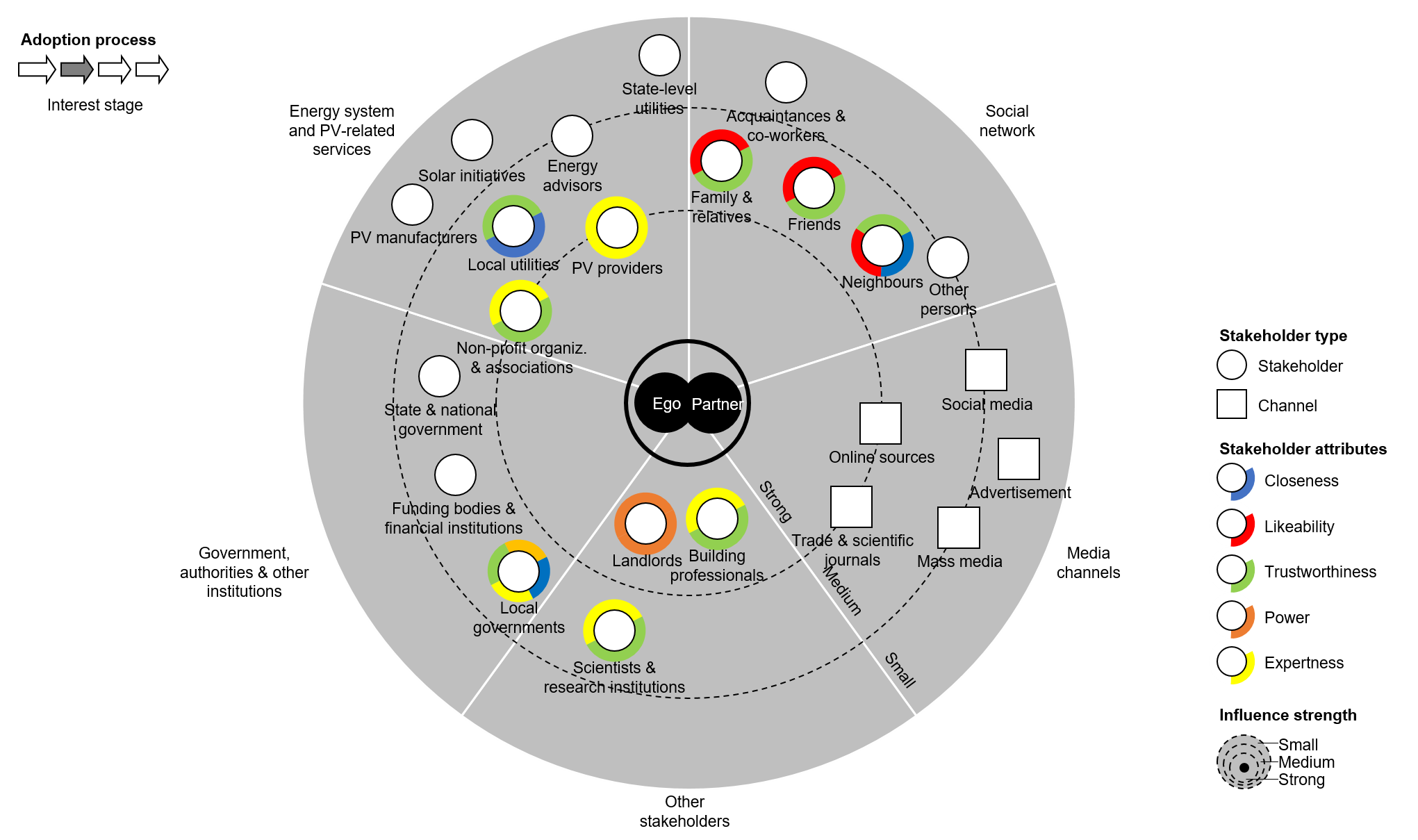}
  \caption{Illustrative overview of egocentric networks of the stakeholder dynamics at the interest stage. The results show the stakeholder dynamics of the focus group discussions in terms of residential PV adoption from the perspective of the household decision-maker (own illustration).}
  \label{fig:EgocentricNetworksInterest}
 \end{figure}

\begin{figure}
  \centering
  \includegraphics[width=1\linewidth]{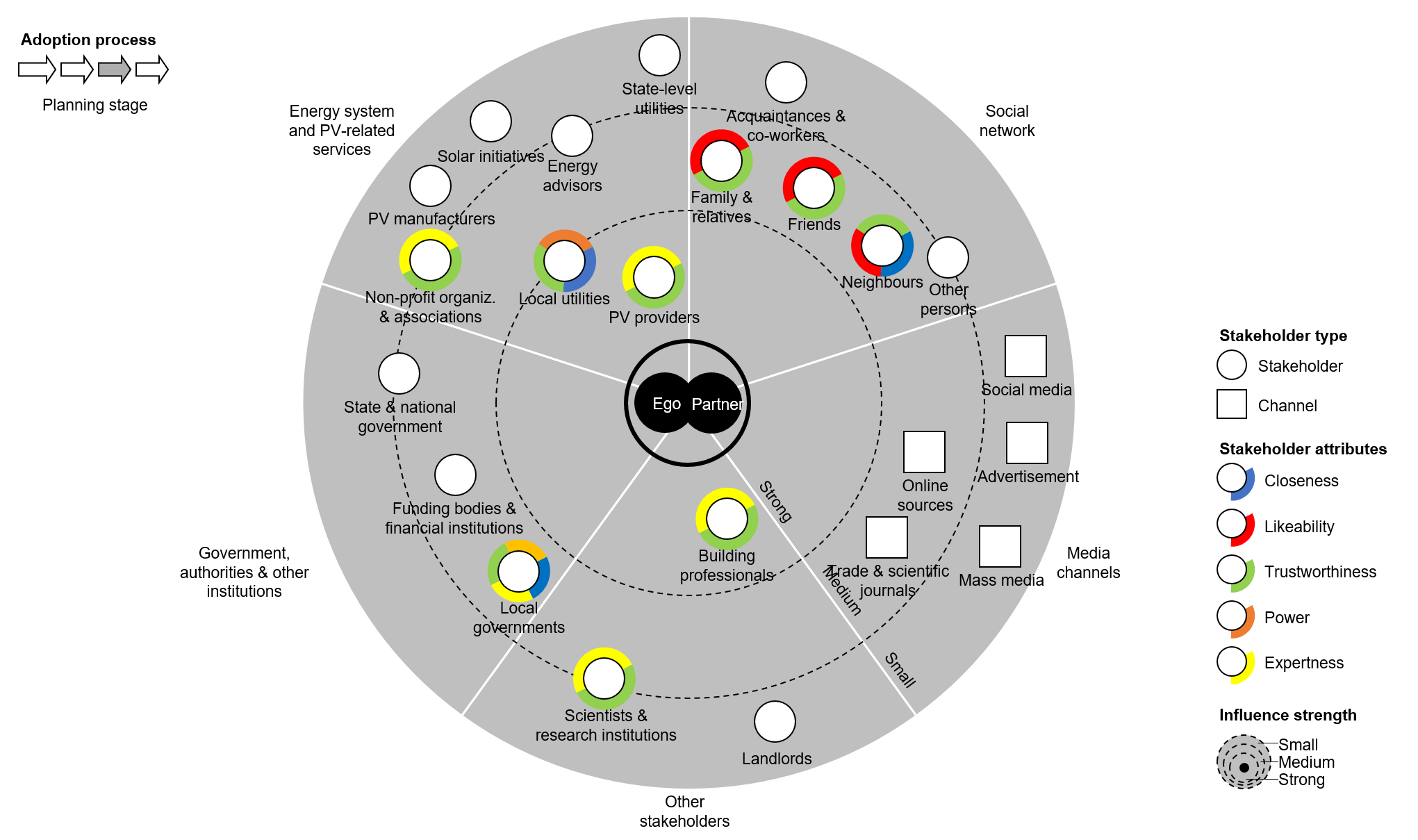}
  \caption{Illustrative overview of egocentric networks of the stakeholder dynamics at the planning stage. The results show the stakeholder dynamics of the focus group discussions in terms of residential PV adoption from the perspective of the household decision-maker (own illustration).}
  \label{fig:EgocentricNetworksPlanning}
 \end{figure}

The stakeholder dynamics visualized as ego network maps of the focus groups for awareness, interest, and planning stages are presented in Figure \ref{fig:EgocentricNetworksAwareness}, \ref{fig:EgocentricNetworksInterest}, \ref{fig:EgocentricNetworksPlanning}. The Tables \ref{focus groupsDs_1}, \ref{focus groupsDs_2},\ref{focus groupsDs_3} in the Appendix give more insights in the assessed category system for each of the decision-stages as raised in Table \ref{tab:category_system}.  
Using network maps allows depicting (a) differences in influence strengths through concentric circles around the decision-maker and (b) perceived stakeholder attributes through actor pies, (c) grouped by stakeholder category through the division into five sectors. A tie is depicted in the map as soon as the ego is influenced in any way in terms of the PV decision by an alter. The stronger a tie, the higher a stakeholder’s influence on the decision-maker and the closer the stakeholder to the decision-maker. The aim is to value the influence strength of one stakeholder in proportion to other stakeholders. The alters encompass all stakeholders identified and analysed. The stakeholder dynamics along the decision process shall be reflected by building a distinct ego network for each process stage. 

Over the stages, out of the 24 stakeholders that have been raised in total, the number of stakeholders with strong influence decreases (8, 7, 5), medium influence initially increases to decrease afterwards (6, 8, 3), and small influence strongly increases in the last decision stage  (10, 9, 16). Whereas the first two stages are characterized by a high number of interactions with diverse actors to raise awareness (14 medium and strong) and accumulate information on PV in general (15 medium and strong), stakeholder interactions in the planning stage are clearly focused on a smaller group (8 medium and strong). In particular, such stakeholders that play a decisive role in the final decision, namely local utilities, PV providers, local governments and authorities and building professionals in the case of retrofitting or new construction, and likeable stakeholders from the social network including partner, family and friends/acquaintances are of importance.\footnote{Because the role of the partner has been identified to be of equally high importance in the stages due to mutual decision-making, it has been decided to associate partner and decision-maker in the centre of the egocentric networks.} Overall, it can be observed that the influence of the social network decreases over the stages, focusing on likeable stakeholders such as friends and family. Apart from the interest stage where neighbours play a central role due to their spatial closeness, their influence is determined by trustworthiness and likeability in the interest and planning stages. PV providers and local utilities as decisive stakeholders for actual adoption of residential PV gain importance during decision-making, whereas no dynamics can be observed with PV manufacturers, state-level utilities and solar initiatives. Although energy advisors and NPOs could both equally serve to consult decision-makers, it is found that the influence of NPOs in the interest stage by far exceeds the influence of energy advisors, which can be explained by the lacking trust in the latter that are alleged to follow primarily commercial interests. Both stakeholders are of the same low importance in the planning stage. Although the local governments and authorities are no implementing stakeholder, the focus groups revealed that their influence increases from low over medium to strong during decision-making. Particularly local governments have been referred to as trustworthy stakeholder. It is assumed that they act in the best interest of the decision-makers. Due to the perceived trustworthiness and independence, and the general idea that politics should more actively work towards a realization of ambitious political goals concerning decarbonization, participants claimed a more central role concerning awareness-raising, information and consultancy. Similarly supporting the idea of a growing desire for reliable, trustworthy information during decision-making, online sources and scientific journals serve as central information sources both in the interest and the planning stage, where decision-makers pro-actively seek reliable information on PV in general, and later on search for expert knowledge to help them make the “right" decision. The role of landlords and building professionals largely depends on the living situation of respondents. Whereas landlords have the power to inhibit PV adoption by their tenants, building professionals acted as change agents during construction or retrofit measures, taking a central role in all three stages.

\section{Discussion of results}
\label{S:5}

\subsection{Assessment of empirical results}
\label{S:5:1}
The results of the focus groups but also of the literature synthesis demonstrate that, indeed, the level of influence of stakeholder categories, stakeholder roles and social dynamics vary among the decision process stages. This can be attributed to the different information requirements in the awareness, interest, and planning stage, where a shift from passive exposure via general to object-specific information procurement takes place. The decision-making process derived from the focus groups aligns with general consensus in scientific literature that with increasing closeness to adoption, considerations become more specific, concrete and context-dependent \cite{Wilson.2018,Arts.2011,Rogers.2003}. Looking into the analysed stakeholder dynamics over the stages, the focus groups showed that the total number of influential stakeholders increases slightly from awareness to interest stage, to strongly decrease in the planning stage, reinforcing the idea of changing involvement of the decision-maker, and changing information requirements. This general understanding is also reflected by the identified role of media which also aligns with findings from literature \cite{Southwell.2014,Reeves.2017,Rai.2016b} and Diffusion of Innovation Theory \cite{Rogers.2003}. Whilst mass media, advertisement and social media play a role only in the awareness stage, where they supply passive decision-makers with information, online sources and scientific journals gain importance in the interest stage, where decision-makers actively gather credible, preferably independent information about PV systems. They lose importance in the planning stage, as they are not able to individually support decision-makers in their planning process. Apart from different media channels raising awareness for PV, \cite{Rai.2016b} showed that direct marketing campaigns of PV providers were “the most popular spark event[s]" to initiate interest in PV, which is reflected by the role of PV providers and advertisement in the awareness stage. However, participants varied in their motivation to autonomously seek information during the interest stage. Whilst some decision-makers pro-actively consulted a large number of information sources, others wait (in vain) for active support and individual advice, explaining that they doubt their ability to find all necessary information for an optimal decision within the plethora of information available. Following \cite{Rogers.2003}, it could be hypothesized that participants vary in terms of novelty seeking and independent judgement-making, leading to passive behaviour among participants less drawn to new (and potentially risky) behaviours. 

The revealed stakeholder dynamics unambiguously show that the influence of the social network changes during the process. The paramount importance of the social network in the awareness stage vanishes, and the more the participants approach the decision, the more emotional closeness, and the less spatial proximity seem to determine influence strength. 
The influence of the partner during the process remains on a high level, as cost-intensive technologies are typically acquired mutually \cite{Verhoog.2017}. In the focus groups, family, friends, neighbours and other private persons are attributed high influence strength in the awareness stage, confirming literature on passive and active peer effects \cite{Bollinger.2012,Graziano.2015,Rode.2016}. The importance of other private persons has received less attention in the literature so far, but appears to be of large importance as visual impressions from unspecified stakeholders such as farmers are as much part of raising awareness towards MOT2 as neighbourhood peer effects. The vanishing role of less emotionally close stakeholders such as neighbours and other private persons beyond initial awareness confirms the assumption of \cite{Palm.2017} that active peer effects are determined by the strength of a pre-established relationship. If for example, a decision-maker sees a PV system on the rooftop of an unfamiliar neighbour or a random farmhouse, this causes an awareness-raising passive peer effect, yet, due to the missing personal relationship, no conversation where information is exchanged (active peer effect) takes place. According to the literature, solar initiatives and civil society groups can exert a positive influence on PV adoption \cite{Noll.2014,Sloot.2018}, particularly in less mature markets \cite{Dewald.2012}. Yet, they have not been identified as important stakeholders in the focus groups. On the one hand, this can be attributed to the limited number of such activities in Leipzig and the state of Saxony, which, according to the experts, is due to limited support from the government. On the other hand, the participants reported strong influences from their social networks, indicating that the diffusion stage where solar initiatives drive adoption is already over. Consequently, experienced peers can be used as trusted, competent information sources, reducing the need to find such sources outside, which taken together confirms the proposition of \cite{Palm.2017,Owen.2014} that the share of previous adopters in the social network conditions the importance of other stakeholders during the later stages of decision-making.

Whereas the search for general information on residential solar PV in the interest stage is characterized by a large number of diverse stakeholder influences, stakeholder dynamics in the planning stage are focused on stakeholders decisive for adoption (local utilities, PV providers, local governments and building professionals in case of construction/renovation) and stakeholders that emotionally support the decision-maker (partner, family, friends/acquaintances).
Whilst this seems plausible, participants reported that they perceived the search for the desired information as difficult, presumably causing a large number of information sources consulted in the interest stage, and hindering a quick identification of relevant stakeholders for the planning stage. Most participants furthermore complain about rapidly changing policies and would value support to simplify information gathering and the planning process. This gap could be initially filled by energy advisors, who could provide independent advice already in the interest stage, support the planning and thereby act as change agents \cite{Owen.2014,Berardi.2013,Michelsen.2012}. However, whilst some participants lacked understanding of the profession, others accused them of commercial interests, jeopardizing a trusting relationship. Instead of relying on economically driven information sources, the overwhelming majority of participants clearly desired a more proactive role of political stakeholders in residential PV adoption. In this context, both local governments and authorities and local utilities have been raised, which is in line with literature that shows the crucial importance of local governments and authorities \cite{Berardi.2013,Jager.2006,Bell.2015} and local utilities \cite{Palm.2016,Reeves.2017,Sommerfeld.2017} in the diffusion of reliable information customized to local needs. Surprisingly, despite participants wish for more active engagement of local utilities, they were not aware of PV and storage products offered directly by the local utility. Furthermore, participants stated that they would value information from NPOs - yet were not aware of the freely available consultancy services provided by e.g. the German Consumer Association or the Saxonian Energy Agency. Interestingly, this discrepancy between the perceived lack of and actually offered services was known to the experts. The apparent mismanagement of reliable information within the plethora of information sources with varying degrees of closeness, likeability, trustworthiness, power and expertness is particularly disadvantageous for individuals that resign more quickly than others, and shift into a passive position, waiting for active support and individual advice. The same can be seen in the context of energy efficiency investments/ measures with incomplete or imperfect information \cite{kangas2018technical,howarth2000economics,palm2018understanding}. As expected, the PV provider plays a central role in the planning stage. However, in search of independent consultancy, some participants reported to have contacted the local governments and authorities to evaluate the product offers received by PV providers, reflecting upon the prevalent wariness of decision-makers to fully trust a commercial stakeholder. The focus groups also reveal differences in stakeholder dynamics depending on the living situation. In the course of new construction and retrofitting projects, building professionals play an important role regarding the transmission of information and evaluation of feasibility. If trusted, they can even be the most influential stakeholder among all stages, either in a positive or a negative way, supporting the findings of \cite{Zedan.2018,Curtius.2018} concerning their gatekeeping influence. In the case of rented properties, landlords can act as highly influential stakeholders in the awareness and interest stage. However, as they have decisive power over the respective rooftop, the decision-making of the tenants ends before the planning stage. Additionally, financial institutions were of moderate influence in the awareness and interest stage, but hardly in the planning stage. This is especially valid for decision-makers who need more financial support than others. Thus, this goes also in line with the literature: Financial considerations are one of the most cited drivers or barriers for PV adoption in the literature \cite{Alipour.2020,Zander.2019}, and participants mentioned local financial institutions to provide support for the initial investment. However, \cite{Huber.2013} showed that existing LCT-related soft loans with less bureaucracy of private banks have not been taken up by citizens as expected, indicating another information mismatch.

An overview of the discussed focus group results with a simultaneous comparison of literature results (cf. \cite{Scheller.2020}) is outlined Table \ref{tab:ComparisonLiteratureEmpirical}. In this context, the table also shows future potential regarding the influence strengths of the individual stakeholders according to the literature and/or the focus groups. Moreover, recommendations for better positioning of the stakeholders are outlined which are generally discussed in the following paragraphs. 

\LTcapwidth=\textwidth
\scriptsize
\begin{landscape}
\begin{longtable}{p{.1\linewidth} p{.045\linewidth} p{.045\linewidth} p{.045\linewidth} p{.045\linewidth} p{.045\linewidth} p{.045\linewidth} p{.05\linewidth} p{.35\linewidth}}
    \caption{Comparison of the results from the literature review (cf. \cite{Scheller.2020}) and the focus group discussions. While the assessment results for each stage demonstrate the actual situation, stated potential is also indicated. Despite the generalization, the policy recommendations must also be viewed with regard to the location of the focus groups (own composition).}\\
    \label{tab:ComparisonLiteratureEmpirical}\\
\toprule
\textbf{Stakeholder} & \multicolumn{3}{p{3.5cm}}{\textbf{Assessment results \newline literature synthesis}}  & \multicolumn{3}{p{3.5cm}}{\textbf{Assessment results \newline focus group discussions}} & {\textbf{Future \newline potential}} & {\textbf{Policy recommendation for an increased PV adoption behaviour of residential decision-makers}}\\
& Stage 1  &   Stage 2 &   Stage 3 & Stage 1  &   Stage 2 &   Stage 3 & & \\
\midrule
\endhead
Family/ relatives & 3 \newline (L,T) & 3\newline (L,T) & 3 \newline (L,T)& 2.5 \newline (L,T) & 2 \newline (L,T)& 2 \newline (L,T) &  &    Sustainable educational strategies for young people to increase the presence of the topic in families  \\
\midrule
Friends & 2,5 \newline (L,T) & 3 \newline (L,T,C) & 3 \newline (L,T,C) & 2,5 \newline (L,T) & 2 \newline (L,T) & 2 \newline (L,T) &  &  Local awareness-raising campaign in which friends are inspired to discuss low-carbon adoption possibilities   \\
\midrule
Acquaintances/ co-workers & 1,5 \newline & 1,5 \newline  & 1,5 \newline & 1,5 \newline & 1 \newline & 1 \newline &  & Presentation of climate protection measures and strategies from individual regional companies   \\
\midrule
Neighbours & 3 \newline (C,T) & 2,5 \newline (C,L,T) & 2,5 \newline (C,L,T) & 3 \newline (C,T) & 2 \newline (C,L,T) & 2 \newline (C,L,T)  && Neighbourhood demonstration events in cooperation with local governments, utilities, or companies  \\
\midrule
Other private \newline persons & 1,5 \newline  & 1,5 \newline & 1,5 \newline & 3 \newline  & 1,5 \newline & 1,5 \newline & &   Public information events at which private individuals share their challenges and solutions during the adoption process  \\
\midrule
Local \newline utilities & 2 \newline (C) & 2 \newline (C,E) & 2,5 \newline (C,E) & 1,5 \newline & 2 \newline (C,T) & 2,5 \newline (C,T,P) & ++ & Promotion of own products by existing customers under consideration of the local conditions and advantages  \\
\midrule
State-level \newline utilities & 1,5 \newline & 1,5 \newline & 1,5 \newline & 1 \newline & 1 \newline & 1 \newline & & More targeted communication strategies regarding the bundled offer with reference to local peculiarities  \\
\midrule
PV \newline manufacturers & 1 \newline & 1 \newline & 1 \newline  & 1 \newline & 1 \newline & 1 \newline & & Proactive technology installation and maintenance advices in regional online forums \\
\midrule
PV \newline providers & 3 \newline (E,C) & 3 \newline (E,T) & 3 \newline (E,T) & 1,5 \newline & 2,5 \newline (E) & 3 \newline (T,E) & & Effective public discussion of the advantages and disadvantages of low-carbon technologies with reference to local peculiarities \\
\midrule
Energy advisors & 1 \newline & 2 \newline (T) & 2 \newline (T) & 1 \newline & 1,5 \newline & 1,5 \newline & + & Positioning as an independent consultant with integrity and active cooperation with local governments and NPOs  \\
\midrule
Non-profit organizations & 1,5 \newline & 2 \newline (T) & 2 \newline (T) & 1 \newline & 2,5 \newline (T,E) & 1,5 \newline (T,E) & & Establishment, support, and announcement of centres for people looking for low-carbon technology adoption advices  \\
\midrule
Solar initiatives/ civil society org. & 3 \newline (C,T) & 3 \newline (C,T) & 3 \newline (C,T) & 1 \newline & 1 \newline& 1 \newline  & + &  Proactive cooperation with citizen groups engaged in adoption activities in terms of statewide or local strategies \\
\midrule
Funding bodies/ financing inst. & 1 \newline & 2 \newline & 3 \newline & 2 \newline & 2 \newline & 2 \newline && Stable range of product orientied soft loans with small bureaucratic efforts    \\
\midrule
Local \newline governments & 2 \newline (T,C) & 2 \newline (T,C) & 2,5 \newline (T,P,C) & 1 \newline & 2 \newline (E,T,P,C) & 2,5 \newline (E,T,P,C) && Role model regarding public buildings and creation multi-stakeholder networks for differing adoption needs  \\
\midrule
State/ national governments & 1,5 \newline & 2 \newline & 2 \newline & 1,5 \newline & 2 \newline & 1,5 \newline && Stable and simple framework conditions with technology-specific funding and less bureaucratic effort   \\
\midrule
Scientists/ \newline research inst. & 1,5 \newline & 1,5 \newline & 1,5 \newline & 1,5 \newline (E,T) & 2 \newline (E,T) & 1,5 \newline (E,T) & + & Local platforms for participation in research and increased communication for the broad population
\\
\midrule
Building \newline professionals & 2 \newline & 1,5 \newline & 1,5 \newline & 2,5 \newline (E,T) & 3 \newline (E,T) & 3 \newline (E,T) & ++ & Advanced sustainable training and appropriate incentive systems for residential adoption decisions  \\
\bottomrule
\multicolumn{9}{l}{Influence strength: [1- small, 2- medium, 3- strong]; Stakeholder attributes: [C- closeness, L- Likeability, T- trustworthy, P- power, E- Expertness]}

\end{longtable}
\end{landscape}
\normalsize	

\subsection{Derivation of practical and theoretical implications}
\label{S:5:2}
The analysis of the focus groups revealed a strong desire among participants for support to successfully navigate through a large amount of information available. This is true for on the one hand the interest stage, where opinions towards PV systems are formed, but also for the planning stage where the decision-maker strives to make the optimal decision concerning the high-cost and long-term investment regarding PV. To determine useful information sources, participants overwhelmingly reported relying on attributes they attach to the stakeholders, typically related to the level of knowledge of the stakeholder, and its relationship with the decision-maker. Stakeholders whose relation is characterized by commercial interests such as energy advisors and PV providers are often viewed with wariness, whereas stakeholders who follow a non-economic agenda such as local governments, NPOs and likeable stakeholders from the social network are perceived as trustworthy, particularly if they act in local proximity.
In order to accelerate residential PV adoption, several implications can be drawn. Responding to the prevalent uncertainty of decision-makers during the information search, the trusted local governments and authorities should exploit their potential and more proactively act as a change agent. In fact, poor engagement of local governments and authorities can even undermine political efforts at higher levels \cite{Berardi.2013}. Despite their technically small role within residential decision-making, they have the potential to become the central stakeholder for decision-makers. In general, it is necessary to accumulate and offer relevant and scientifically based information on PV systems. Besides, guidance within the planning process by either offering consultancy services itself, or redirecting to independent, cost-free services offered by stakeholders such as the German Consumer Association or the Saxonian Energy Agency needs to be given. Moreover, local governments and authorities could act as a role model and raise awareness themselves, e.g. by equipping municipal buildings with rooftop PV. 
Beyond the local governments and authorities directly engaging with decision-makers, they could initiate and coordinate a multi-stakeholder network on the local level, including actors such as the local utilities, PV providers, and NPOs, but also building professionals. Such a multi-stakeholder approach has proven very fruitful in the city of Bottrop, as joining forces is useful to develop and put into place a strategic approach to push low-carbon technology adoption \cite{mattes2015energy,best2019energiewende}. Initially, the local stage of diffusion should be assessed, and in case of low diffusion rates, solar initiatives or related interest groups acting as change agents should be initiated and supported in situations where peer effects are not effective. Furthermore, stakeholders involved in construction or retrofit processes such as building professionals can approach decision-makers at the right time and redirect them to the information pool of the local governments and authorities, ensuring in particular that passive decision-makers are pushed towards MOT2, and receive support in opinion-making and planning. Besides, information events \cite{Jager.2006} or incentive programs such as 100 rooftops or recruit a friend could be launched in cooperation of local governments and authorities, utilities and PV providers, utilizing local newspapers as a media channel, and benefiting from the local proximity to the decision-makers. Overall, a proactive, central role of the local governments and authorities is crucial in accelerating local adoption rates, as it could provide non-commercially oriented guidance for decision-makers, and help to overcome information barriers resulting in a discrepancy between perceived and actually offered services.

\section{Concluding remarks}
\label{S:6}

While the insights from the literature indicated a multitude of different stakeholders that might influence the PV-related decision-making process in a dynamic way (Section \ref{S:2.2}), we could not find a comprehensive study that analyses and assesses the influences of the mentioned stakeholders of different social groups on residential decision-makers with respect to PV. The individual pieces of information had to be put together with a view of the various decision-making phases. Thus, the empirical findings of the focus groups as outlined in Sections \ref{S:4.4} and \ref{S:4.5} represent a novel investigation with a certain focus on the different stages in a complex decision environment. The four focus groups made it possible to match but also deepen knowledge. Finally, the comparison and synthesis of the results of the focus groups with the existing literature body (Sections \ref{S:5:1}) enabled the derivation of valid implications (Sections \ref{S:5:2}).

Indeed, the results demonstrate that stakeholder influences are dynamic on several levels. At the awareness stage, where predominantly passive exposure raises awareness among decision-makers, proximity and ordinary communication on a regular basis play a central role. Trustworthiness and expertness are of importance both in the interest and the planning stage. However, as during planning, the decision-maker must rely on commercial stakeholders, the attribute trustworthiness seems to become less important. Yet, the strong desire of decision-makers for trustworthy, independent information sources and consultancy draws attention to the unexploited potentials of stakeholders who unify such attributes: local governments and authorities as well as NPOs, with local governments having an even larger potential. This is due to the fact that local governments and authorities are perceived as legitimate and credible actors in LCT decision-making. Household decision-makers even claim local public stakeholders to assume responsibility in achieving climate goals. The local governments and authorities should therefore seize their position to accelerate residential PV adoption by directly guiding the decision process, for example, by implementing a local multi-stakeholder network.

As methodological choices never come without a trade-off, several limitations can be found regarding the results presented in this article: 
In terms of the empirical design, we see three limitations that all affect the representativeness of our results. First, we prioritized quality over quantity by conducting only four focus groups. Due to budget and time constraints, this low number was chosen to still allow the research team to focus on the in-depth evaluation of the focus groups over conducting as many focus groups as possible (as suggested by, e.g., \cite{GAILING2018355}). Second, our sample consisted of participants stemming from milieus likely to be able to adopt PV, thereby excluding a large part of society. Third, while the stakeholder landscape was explored comprehensively, not all participants stated their perceptions of every single stakeholder, limiting the available frequency data on influence strength. At the same time, the synthesis of the focus group results with the literature findings allowed a certain representative evaluation in a broader context.

Further limitations are posed by the chosen methodology. While the combination of focus groups and content analysis are seen as rather rigorous methods within qualitative research \citep{SOVACOOL201812}, they are by their nature interpretive and subjective. To mitigate this limitation, the research design used a stakeholder tableau as a framework and the content analysis was performed by two researchers to arrive at an inter-subjective interpretation. 
Also, method-inherent limitations were carried over into our research; as such, participants might have been influenced by the salience of the research agenda and their responses could have been sub-consciously manipulated \cite{SAGE.SocialResearchMethods}. Furthermore, the researchers were aware that their interactions within this formal setting differed from interactions in a more natural setting. While we believe that this had an influence on the finer details of the research, the more salient insights and the relative importance of the different stakeholders should broadly remain valid.

In terms of generalizability, we see severe limitations of our results regarding validity for other low-carbon technologies. Nevertheless, similar to PV systems,e.g., electric vehicles represent technical but also visible products with high upfront, and low operation cost, leading to a positive long-term effect \cite{Rezvani.2015, Sierzchula.2014}. The presented stakeholder structure and influences can at least be transferred to technologies with similar traits. Furthermore, the results of the analysis are rather specific both in geographical and in product scope. While the institutional infrastructure and political framework differs in other countries from the investigated case, the literature search has revealed that there are several studies on PV and also LCT decision-making in general, for instance in the United States and Sweden that provide valuable insights for this research. Thus, our findings might be also transferable when the framework conditions are broadly similar.

While being a fundamental limitation for every research, researcher and participant bias is particularly problematic in contexts where the issue of investigation is constructed as a socially negotiated response. Whereas the objective of this research was precisely to investigate the (inter-)subjective construction of the perceived and negotiated stakeholder context of the investigated participants, these results need to interpreted with possible bias and the social dynamics found within the groups in mind, impeding its generality. To mitigate bias, the research team based the investigation on a theoretical framework and broad literature basis, employed an experienced focus group moderator from a third party not involved in the theoretical conception of the analytical framework (and thus less prone to the respective bias) and performed the content analysis independently between two researchers. While this mitigated some of the bias, the reader should be aware of the residual bias in contextualizing our results.

In order to address the limitations raised and to establish a more profound understanding, future research needs to verify the stated stakeholder dynamics while considering the underlying principles such as the stakeholder attributes. The goal is an operationalization of hypotheses with, e.g., representative quantitative surveys. In our future research, we are aiming to carry out exactly such quantitative analyses for a sub-area, namely energy system models for predicting PV adoptions on the basis of empirically grounded agent-based modelling \cite{johanning2020modular,Scheller.2019}.

\section*{Supplementary material} 
The Supplementary Material consists of the discussion guide for the regular focus groups 1, 2 and 3 (SM A), the discussion guideline for focus group 4 (SM B), description of the stakeholders and stakeholder categories (SM C), brief profiles of selected Sinus-Milieus\textsuperscript{\textregistered} (SM D), a summary of practical implementations of focus groups (SM E), characteristic quotes of focus groups regarding the conceptual understanding (SM F), characteristic quotes of focus groups regarding the understanding of the decision process (SM G), characteristic quotes of focus groups regarding the stakeholder dynamics during the decision-making process (SM H). In this context, the original quotations of the participants of the focus groups in German and the English translations are presented concerning the process stage and the respective stakeholder. The complete and anonymous transcripts of the focus groups 1-4 can be obtained in the original German language on request.

\section*{Declarations of interest} 
There are no competing interests.

\section*{Acknowledgement}
The authors wish to thank Sinus Markt- und Sozialforschung GmbH and especially the project team members Jochen Resch, Silke Borgstedt, and James Rhys Edwards for the cooperation in organizing and conducting the focus group meetings. Fabian Scheller, Emily Schulte, Isabel Doser, Simon Johanning receive funding from the project SUSIC (Smart Utilities and Sustainable Infrastructure Change) with the project number 1722 0710. This study is financed by the Saxon State government out of the State budget approved by the Saxon State Parliament. Fabian Scheller also kindly acknowledges the financial support of the European Union's Horizon 2020 research and innovation programme under the Marie Sklodowska-Curie grant agreement no. 713683 (COFUNDfellowsDTU).

\appendix

\section{Overview stakeholder descriptions for focus group discussions}

\begin{table}[H]
\footnotesize
\renewcommand{\arraystretch}{1.2}
\setlength{\tabcolsep}{5pt}
    \caption{Stakeholder mapping and classification based on their role in the PV adoption processes (adopted by \cite{Scheller.2020}).}
    \label{tab:stakeholder}
\begin{tabularx}{\textwidth}{p{0.2\textwidth} p{0.75\textwidth}}
\toprule
\textbf{Stakeholder} & \textbf{Description} \\ 
\midrule
\multicolumn{2}{l}{\textbf{Category: Social network}}\\
\hline
Partner & Spouse or life partner with whom decision is made \\
\hline
Family & Family members and relatives  \\
\hline
Friends & Persons in decision-maker’s (immediate) social circle  \\
\hline
Acquaintances/ \newline co-workers & Persons in decision-maker’s wider social circle  \\
\hline
Neighbours & Persons living in decision-maker’s neighbourhood \\
\hline
Other private persons & Private persons outside decision-maker’s social circle with an interest in photovoltaic (e.g., fair visitors, forum contributors) or other previous adopters  \\
\midrule
\multicolumn{2}{l}{\textbf{Category: Energy system and PV-related services}}\\
\hline
Local utilities & Energy providers operating locally (e.g., city or communal providers) \\
\hline
State/ national-level utilities & Energy providers operating in many locations (e.g., E.On, RWE)  \\
\hline
PV manufacturers & Photovoltaic producers (if a PV manufacturer also sells PV systems to customers at household level, it is considered as PV provider)  \\
\hline
PV providers & Private companies that sell and install PV systems to household decision-makers, often providing additional services such as consulting and maintenance   \\
\hline
Energy advisors & Experts in the private sector who offer information or advice regarding energy issues for a fee \\
\hline
Non-profit organizations/ associations & Independent (public) organizations that offer information or advice regarding energy issues (e.g., consumer organizations, energy agencies, non-profit organizations) mostly free of charge  \\
\hline
Solar initiatives & Formal or informal organizations stemming from civil society engaged in the support of renewable energies and PV in particular \\
\midrule
\multicolumn{2}{l}{\textbf{Category: Government/authorities and other institutions}}\\
\hline
Funding bodies/ financial institutions & Financial institutions that play a role in financing photovoltaic investments (e.g., banks, development banks, credit unions)  \\
\hline
Local governments & Communal policy makers, government offices and organizations (e.g., city hall, communal building authority)  \\
\hline
State / national governments & Policy makers and authorities at state and national level
 \\
\midrule
\multicolumn{2}{l}{\textbf{Category: Others}}\\
\hline
Building professionals & Professionals in the construction sector including private companies that plan and undertake building construction and renovation work (e.g., builders, architects)
  \\
  \hline
Landlords & Associations or individuals owning living space (relevant for decision-makers that are tenants)
  \\
\hline
Scientists & Researchers engaged in research on energy-related issues (e.g., from research institutions or universities) \\
\midrule
\multicolumn{2}{l}{\textbf{Category: Media channels*}}\\
\hline
Advertisement & All kinds of off- and online advertisement (e.g., leaflets, banners) united by promotional intention \\
\hline
Online sources & Unspecified online sources (e.g., “googling around”)  \\
\hline
Social media & Online communities (e.g., Facebook, Instagram, Twitter, YouTube) \\
\hline
Mass media & Off- and online media coverage in newspaper, magazines and television  \\
\hline
Trade/ scientific journals & Energy- or technology-related (scientific) journals (e.g., Photon)
 \\
\bottomrule
\multicolumn{2}{p{16cm}}{\textsuperscript{*} Media plays a particular role since the channels are not interpreted as proper stakeholders. The channels are used by all different kind of stakeholders. In this context, the tangible stakeholders “behind” seem to be rapporteurs who are responsible for media coverage but they cannot be treated like the other mentioned actors. Instead of directly integrating rapporteurs as actors into the decision-making landscape, the channels through which information is conveyed, are taken into account. Usually, rapporteurs don’t act for their own sake but on behalf of certain stakeholders to get information across society as target group. In this regard, it is not always visible which stakeholder really is behind media coverage.}
\end{tabularx}
\end{table}

\section{Frequency count of stakeholder interactions in the different decision-making stages}
\begin{table}[H]
\footnotesize
\renewcommand{\arraystretch}{1.2}
\setlength{\tabcolsep}{5pt}
\caption{Focus groups on the interaction of stakeholders (alter) with the household decision-maker (ego) in the awareness stage. While entries of the stakeholder list are related to the mentions in all decision-stages, the frequency count for each discussion (focus groups 1-3) and in total (N), and the influence strength is related to the stage indicated. The final assessment takes the stakeholder attributes as soft indicators into account (own composition).}
\label{focus groupsDs_1}
 
\begin{tabularx}{\textwidth}{p{.2\textwidth} p{.06\textwidth} p{.06\textwidth} p{.06\textwidth} p{.02\textwidth} p{.3\textwidth} p{.25\textwidth}}

\toprule
\textbf{Stakeholder} & \multicolumn{3}{p{3.5cm}}{\textbf{Frequency count per \newline focus group discussion}} & \textbf{N} & \textbf{Further criteria determining\newline strength of influence} & \textbf{Final assessment of \newline strength of influence}\\
 & Focus group 1 & Focus group 2 & Focus group 3 & & & \\

\hline
\hline

Partner & X(2) & X(1) & X(1) &4 & Likeability/trustworthiness/closeness & Strong \\
\hline
Family & & X(2) & & 2&Likeability/trustworthiness & Strong\\
\hline
Friends/acquaintances & X(2) & X(1) & X(4) & 7& Likeability/trustworthiness & Strong\\
\hline
Co-workers& X(1) & & X(1) & 2 & & Small/none \\
\hline
Neighbours& X(3) & X(3) & X(1) & 7 & Closeness/trustworthiness & Strong \\ 
\hline
Other private persons & X(4) & X(2) & X(1) & 7 & & Strong \\
\hline
Local utilities & & & & & & Small/none\\
\hline
National utilities & & & & & & Small/none\\
\hline
PV manufacturers & & & & & & Small/none\\
\hline
PV providers & & & X(1) & 1 & & Small/none\\
\hline
Energy advisors & & & & & & Small/none\\
\hline
NPOs/associations & & & & & & Small/none\\
\hline
Solar initiatives & & & & & & Small/none\\
\hline
Funding bodies/\newline financial institutions & X(2) & X(1) & X(1) & 4 & & Middle\\
\hline
State and \newline national govern\-ments& X(2) & X(1) & & 3 & & Middle\\
\hline
Local governments& & X(1) & & 1 & & Small/none\\
\hline
Building professionals & & X(1) & & 1 & Expertness/trustworthiness&Strong \\
\hline
Landlords & X(3) & & & 3 & Power (ownership) & Strong\\
\hline
Scientists & X(1) & X(1) & X(1) & 3 & Expertness/trustworthiness& Middle\\
\hline
Advertisement & X(2) & X(1) & & 3 & & Middle\\
\hline
Online sources& X(6) & & X(1) & 7 & & Middle\\
\hline
Social media & X(1) & X(2) & & 3 & & Strong\\
\hline
Mass media & X(2) & X(2) & X(3) & 7 & & Middle\\
\hline
Scientific journals & & & & & & Small/none\\
\bottomrule
\end{tabularx}

\end{table}

\begin{table}[H]
 \footnotesize
\renewcommand{\arraystretch}{1.2}
\setlength{\tabcolsep}{5pt}
\caption{Focus groups on the interaction of stakeholders (alter) with the household decision-maker (ego) in the interest stage. While entries of the stakeholder list are related to the mentions in all decision-stages, the frequency count for each discussion (focus groups 1-3) and in total (N), and the influence strength is related to the stage indicated. The final assessment takes the stakeholder attributes as soft indicators into account (own composition).}
\label{focus groupsDs_2}

\begin{tabularx}{\textwidth}{p{.2\textwidth} p{.06\textwidth} p{.06\textwidth} p{.06\textwidth} p{.02\textwidth} p{.3\textwidth} p{.25\textwidth}}

\toprule
\textbf{Stakeholder} & \multicolumn{3}{p{3.5cm}}{\textbf{Frequency count per \newline focus group discussion}} & \textbf{N} & \textbf{Further criteria determining\newline strength of influence} & \textbf{Final assessment of \newline strength of influence}\\
& Focus group 1 & Focus group 2 & Focus group 3 & & & \\
\hline
\hline

Partner & X(1) & X(3) & X(1) &5 & Power (mutual decision) & Strong \\
\hline
Family & & X(1) & & 1&Likeability/trustworthiness & Middle\\
\hline
Friends & X(1) & X(1) & X(1) & 3& Likeability/trustworthiness & Middle\\
\hline
Acquaintances/ & & X(3) & & 3 & & Small/none \\
\hline
co-workers & & & & & & \\
\hline
Neighbours & X(1) & X(2) & & 3 & Closeness/trustworthiness/likeility & Middle \\ 
\hline
Other private persons & & & & & & Small/none \\
\hline
Local utilities & X(1) & X(3) & & 4 & Closeness/trustworthiness & Middle\\
\hline
National utilities & X(1) & & & 1 & & Small/none\\
\hline
PV manufacturers & X(1) & X(2) & & 3& & Small/none\\
\hline
PV providers & X(2) & X(5) & X(4) & 11 & Expertness & Strong\\
\hline
Energy advisors & & & & & & Small/none\\
\hline
NPOs/associations & & X(1) & X(1) & 2 & Expertness/trustworthiness & Middle\\
\hline
Solar initiatives & & & & & & Small/none\\
\hline
Funding bodies/\newline financial institutions & & X(2) & X(3) & 5 & & Middle\\
\hline
State and \newline national govern\-ments& X(2) & X(1) & X(2) & 5 & & Middle\\
\hline
Local governments& X(3)& X(1) & X(1) & 5 & Closeness/trustworthiness/\newline expertness/power & Strong\\
\hline
Building professionals & X(2) & X(2) & & 4 & Expertness/trustworthiness&Strong \\
\hline
Landlords & & X(1) & & 1 & Power (ownership) & Strong\\
\hline
Scientists & X(1) & X(1) & X(1) & 3 & Expertness/trustworthiness& Middle\\
\hline
Advertisement & & & & & & Small/none\\
\hline
Online sources & X(6) & X(2) & X(2) & 10 & & Strong\\
\hline
Social media & & X(1) & & 1 & & Small/none\\
\hline
Mass media & & X(3) & & 3 & & Small/none\\
\hline
Scientific journals & X(1) & X(5) & X(1) & 7& Expertness/trustworthiness & Strong\\
\bottomrule

\end{tabularx}

\end{table}

\begin{table}[H]
 \footnotesize
\renewcommand{\arraystretch}{1.2}
\setlength{\tabcolsep}{5pt}
\caption{Focus groups on the interaction of stakeholders (alter) with the household decision-maker (ego) in the planning stage. While entries of the stakeholder list are related to the mentions in all decision-stages, the frequency count for each discussion (focus groups 1-3) and in total (N), and the influence strength is related to the stage indicated (own composition). The final assessment takes the stakeholder attributes as soft indicators into account (own composition).}
\label{focus groupsDs_3}
 
\begin{tabularx}{\textwidth}{p{.2\textwidth} >{\columncolor[gray]{.8}} p{.06\textwidth} p{.06\textwidth} p{.06\textwidth} p{.02\textwidth} p{.3\textwidth} p{.25\textwidth}}

\toprule
\textbf{Stakeholder} & \multicolumn{3}{p{3.5cm}}{\textbf{Frequency count per \newline focus group discussion}} & \textbf{N} & \textbf{Further criteria determining\newline strength of influence} & \textbf{Final assessment of \newline strength of influence}\\
 & Focus group 1 & Focus group 2 & Focus group 3 & & & \\
\hline
\hline

Partner & & & X(2) &2 & Power (mutual decision) & Strong \\
\hline
Family & & & & & & Middle\\
\hline
Friends/acquaintances & & X(1) & X(3) & 4& & Middle\\
\hline
Co-workers& & & X(2) & 2 & & Small/none \\
\hline
Neighbours& & & X(1) & 1 & & Small/none \\ 
\hline
Other private persons & & & X(1) & 1 & & Small/none \\
\hline
Local utilities & & X(2) & X(1) &4 & & Strong\\
\hline
National utilities & & & X(1) & 2& & Small/none\\
\hline
PV manufacturers & & & X(1) &1 & & Small/none\\
\hline
PV providers & & X(4) & X(2) & 6 & Expertness/trustworthiness & Strong\\
\hline
Energy advisors & & & & & & Small/none\\
\hline
NPOs/associations & & & X(1) & 1& & Small/none\\
\hline
Solar initiatives & & & & & & Small/none\\
\hline
Funding bodies/\newline financial institutions & & & & & & Small/none\\
\hline
State and \newline national govern\-ments& & & & & & Small/none\\
\hline
Local governments& & &X(1) & 1 & Power& Strong\\
\hline
Building professionals & & X(1) & & 1 & Expertness/trustworthiness&Strong \\
\hline
Landlords & & & & & & Small/none\\
\hline
Scientists & & & X(1) & 1 & &Small/none\\
\hline
Advertisement & & & & & & Small/none\\
\hline
Online sources & & & X(1) & 1 & & Middle\\
\hline
Social media & & & & & & Small/none\\
\hline
Mass media &&&&&&Small/none\\
\hline
Scientific journals&&&&&&Small/none\\
\bottomrule
\multicolumn{6}{l}{\textsuperscript{*} focus groups 1 neglected the third stage due to the low share of previous adopters}
\end{tabularx}

\end{table}


\newpage
\bibliographystyle{elsarticle-num}
\bibliography{ModelDescription.bib}







\end{document}